\documentclass[12pt,reqno]{amsart}
\usepackage{natbib}
\usepackage{graphics,graphicx,amsmath,amssymb,epsfig,stmaryrd,color,flafter,xspace,lscape,threeparttable,pdfpages}
\usepackage{csquotes,amsfonts,amsthm,geometry,array,booktabs,latexsym,caption,subfig,multirow,tabularx,rotating,bigstrut}
\usepackage[font=small,labelfont=bf]{caption}
\DeclareCaptionFont{tiny}{\tiny}
\everymath{\displaystyle}

\RequirePackage{setspace,indentfirst,epsfig,psfrag,ifthen,ifpdf}
\usepackage{changepage}

\theoremstyle{plain}

\newtheorem{assumption}{Assumption}

\newtheorem{subassumption}{Assumption\normalfont}[assumption]

\newenvironment{assumption*}
 {\ifnum\value{subassumption}=0 \stepcounter{assumption}\fi\subassumption}
 {\endsubassumption}
\newenvironment{assumption+}[1]
 {\renewcommand{\thesubassumption}{#1}\subassumption}
 {\endsubassumption}

\newcommand\blfootnote[1]{%
  \begingroup
  \renewcommand\thefootnote{}\footnote{#1}%
  \addtocounter{footnote}{-1}%
  \endgroup
}

\theoremstyle{definition}
\newtheorem{assump}{Assumption}

\newtheorem{definition}{Definition}
\newtheorem*{definition*}{Definition}

\newtheorem{lemma}{Lemma}

\newtheorem{proposition}{Proposition}
\newtheorem*{proposition*}{Proposition}

\newtheorem*{remark*}{Remark}

\newtheorem{fact}{Fact}

\numberwithin{equation}{section}

\begin{document}

\title[Roy model and occupational segregation]{Occupational segregation in a Roy model \\ with composition preferences}
\author{Haoning Chen}
\author{Miaomiao Dong}
\author{Marc Henry}
\author{Ivan Sidorov}
\noindent \blfootnote{\scriptsize{The first version is of October 15, 2020. The present version is of \today. 
The authors gratefully acknowledge helpful comments and suggestions from the associate editor and a referee, as well as helpful discussions with Kalyan Chatterjee, Lena Edlund and Isma\"{e}l Mourifi\'e. Correspondance address: Marc Henry, 522 Kern, Department of Economics, Penn State, University Park, PA 16802. E-mail: marc.henry@psu.edu.}}

{\scriptsize 

\begin{abstract}
We propose a model of labor market sector self-selection that combines comparative advantage, as in the Roy model, and sector composition preference. Two groups choose between two sectors based on heterogeneous potential incomes and group compositions in each sector. Potential incomes incorporate group specific human capital accumulation and wage discrimination. Composition preferences are interpreted as reflecting group specific amenity preferences as well as homophily and aversion to minority status. We show that occupational segregation is amplified by the composition preferences and we highlight a resulting tension between redistribution and diversity. The model also exhibits tipping from extreme compositions to more balanced ones. Tipping occurs when a small nudge, associated with affirmative action, pushes the system to a very different equilibrium, and when the set of equilibria changes abruptly when a parameter governing the relative importance of pecuniary and composition preferences crosses a threshold.
\end{abstract}

}

\maketitle
\thispagestyle{empty}

{\scriptsize 

}




\section*{introduction}

Occupational segregation, particularly the under-representation of women in mathematics intensive (hereafter STEM) fields is well documented (see for instance \cite{KG:2017}). The under-representation of women in STEM fields is an important driver of the gender wage gap (see \cite{daymont1984}, \cite{zafar2013} and \cite{SHB:2019}, \citet{BK:2017}). Insofar as it reflects talent misallocation, it has significant growth implications (see \citet{Baumol:90}, \citet{MSV:91}, and \citet{HHJK:2019}). Occupational segregation may also affect welfare directly, through composition effects in utility (see \citet{FSL:2020}).

Traditional explanations of the self-selection of women out of STEM fields hinge on pecuniary considerations.
Differences in labor force participation profiles are blamed. According to \citet{Zellner:75}, women weigh early earnings more than career progression. \citet{MP:74} and \citet{Polachek:81} explain differences in human capital accumulation over the life cycle by household specialization. \citet{Becker:85} ascribes occupational segregation to differences in productivity, due to differential household responsibilities. 
Differences in productivity may also arise as a result of 
differences in field specific human capital accumulation as a result of gender stereotyping of learning and occupations, (see \citet{ES:2010} and \citet{Carlana:2019}). \citet{Francois:98} explains occupational segregation with multiple equilibria arising from the interaction between household and labor market specializations. Finally, incomes of men and women may differ as a result of wage discrimination in STEM fields (see for instance \citet{BCJW:2016}), or from weak prejudice in wage postings as in \citet{LMD:2005}.

Beyond pecuniary considerations, selection into STEM may be driven by composition preferences, either directly, through homophily (see \citet{Arrow:98}) or aversion to being in a small minority, or indirectly, through amenity provision. Minority stress is documented in \citet{Meyer:95}. Sexual harassment is more prevalent in male dominated occupations (\citet{GM:82} and \citet{WSL:2007}), and is measured in terms of compensating differentials in \cite{Hersch:2011}. \citet{Shan:2020} shows evidence of increased attrition of women randomly assigned to male dominated groups in experimental settings and \citet{BW:2021}  shows evidence of increased attrition of women in graduate cohorts with no female peers. Composition preferences may also reflect preferences for amenities, such as workplace flexibility, whose provision depends on gender composition (see \citet{Usui:2008}, \citet{Glauber:2011}, \citet{WZ:2018} and \citet{MP:2017}).

The model we propose here incorporates both types of forces: monetary and composition preferences. Individuals from two groups (e.g., women and men) choose their sector of activity based on their draw from a joint distribution of potential incomes (which incorporate all pecuniary incentives described above) and a preference for composition (that reflects group specific amenity provision, homophily and aversion to minority status). Potential incomes are heterogenous within and between groups, whereas composition preferences are heterogeneous between groups only. Our model therefore combines the Roy model of self-selection, in \citet{Roy:51}, with models where behavior is affected by a group average, as in \citet{CJ:88}, \citet{Bernheim:94} and \citet{BD:2001} or compositions, as in \citeauthor{Schelling:71} [\citeyear{Schelling:69,Schelling:71,Schelling:73,Schelling:78}].\footnote{As a mean-field discrete choice game, our model also resembles the \citet{EF:2003} model of agglomeration, with two groups but no heterogeneity within group, and \citet{Blonski:99}, with binary actions and heterogeneous preferences, but a single group.}

We uncover two types of phenomenon associated with the interaction of comparative advantage and composition preference: amplification of occupational segregation and tipping. We show that composition preferences tend to amplify occupational segregation relative to compositions induced by comparative advantage only, i.e., compositions that would arise in a Roy model of sector self-selection without composition preferences. The scale of the amplification effect depends on the relative importance of comparative advantage, on the one hand, and the intensity of composition preferences, on the other. We propose parameterizations of the distribution of comparative advantage and composition preference functions that allow us to quantify the scale of the amplification effect. 

The combination of comparative advantage and composition preferences also produces tipping phenomena. \citet{Pan:2015} proposes a model of tipping in labor markets.  In \citet{Pan:2015}\footnote{The tipping model in \citet{Pan:2015} is equivalent to the model in Card et al. (2008), except that the inverse housing demand curves are replaced with inverse labor supply curves.}, there is a single market clearing wage (no discrimination) which determines equilibrium, and the inverse labor supply curve of men depends on the proportion of women in the occupation. As the labor supply of women increases exogenously, there comes a point where the three market equilibria (on stable and one unstable with mixed labor force and one stable with women only) turn into two (the stable mixed and women only equilibria) and finally into one (the women only equilibrium). Hence there is tipping from the stable mixed gender labor force equilibrium to the women only labor force equilibrium. This is tipping to full segregation, as in the original Schelling model.

In contrast with \citet{Pan:2015}, there is heterogeneity in potential incomes in our model, so that comparative advantage competes with composition preferences to determine sector selection, and tipping can occur from extreme to more balanced compositions. We document two types of tipping. First, we call {\em nudge tipping} a shift between two very different equilibria as a result of a very small nudge (see \citet{HK:2006}). Second, we call {\em catastrophic tipping} a sudden change in the set of equilibria, and the disappearance of an equilibrium with extreme compositions as parameters cross a threshold (which corresponds to the notion of tipping point in catastrophe theory).
We associate the former with quotas (see \citet{BBJL:2019}), and the latter with regulatory changes in workplace culture.

As is well known (see \citet{HH:90} and references therein), estimation of Roy models presents challenges, as the distribution of primitive potential incomes~$(Y_1,Y_2)$ cannot be recovered from observed realized incomes and sector choice. \citet{MHM:2020} derive sharp bounds on the distribution of potential incomes in a model of sector selection based on incomes only. In the Roy model with composition preferences, we derive sharp bounds on the joint distribution of potential incomes and composition preference functions based on the observation of realized compositions and distributions of realized outcomes by sector and by group. Within the parametric specification of comparative advantage and composition preference functions, this allows us to conduct inference on the relative intensity of composition preferences and pecuniary considerations using existing moment inequality inference procedures.




\section{Roy model with composition preferences}
\label{sec:frame}


\subsection{Description of the model}
\label{sec:des}

We consider two groups of individuals,~$w$ and~$m$. There are exogenous masses~$\mu_w$ of group~$w$ individuals and~$\mu_m$ of group~$m$ individuals. All individuals simultaneously choose between two sectors of activity: sectors~$1$ and~$2$.
Individuals within each group have heterogeneous potential incomes in each sector, which are exogenously determined. Let~$(Y_{1t},Y_{2t})$ be the pair of potential incomes of an individual randomly drawn from the group~$t$ sub population,~$t\in\{w,m\}$. This group~$t$ individual knows that they will receive income~$Y_{1t}$ (resp.~$Y_{2t}$) if they choose sector~$1$ (resp.~$2$). This group~$t$ individual chooses sector~$1$ (resp.~$2$) when
$Y_{1t}-h_{t}\left(\mu_{1t}/\mu_1\right)$ is larger (resp. smaller) than~$Y_{2t}-h_{t}\left(\mu_{2t}/\mu_2\right)$, where~$\mu_{jt}/\mu_j$ is the ratio of the endogenous mass~$\mu_{jt}$ of group~$t$ individuals employed in sector~$j$ over the endogenous total mass~$\mu_j$ of individuals employed in sector~$j$, and~$h_t$ is a decreasing function. We make the following regularity assumption on the primitives of the model.
\begin{assumption}
\label{ass:reg}
The function $h_t$, $t\in\{w,m\}$, is
continuously differentiable. The cumulative distribution~$F_{\Delta_t}$ of sector~1 advantage~$\Delta_t=Y_{1t}-Y_{2t},$ $t\in\{w,m\}$, is continuously differentiable with a continuous inverse $F_{\Delta_t}^{-1}$, and~$\mathbb P[\Delta_t>0]\in(0,1)$. 
\end{assumption}

The model described in section~\ref{sec:des} treats the distribution of the vector of potential incomes~$(Y_{1t},Y_{2t})$ of an individual~$i$ of group~$t$ as a primitive. Potential incomes are defined as the incomes individual~$i$ would command in each sector of activity. The potential incomes are heterogeneous at the individual level. They incorporate talent, general and sector specific human capital accumulation and their price on the labor market. The distributions of potential incomes are group-specific. Different groups may have accumulated different amounts of general and sector specific human capital because of differential access to learning, possibly due to group profiling in society. Different groups may also face different prices on the labor market because of wage discrimination.
The non-pecuniary component in preferences depends only on group and sector group compositions, so that, unlike income, there is only group-level heterogeneity in composition preferences. The functions~$h_t$,~$t\in\{w,m\}$, reflect composition preferences. They incorporate pure social interaction effects, where homophily can be modeled by non increasing~$h$ and aversion to being in a small minority can be modeled with large values of~$h(u)$, when~$u$ is close to~$0$. Composition preferences also reflect group-specific preferences for amenities, whose provision depends on composition. Amenities include flexibility of working hours, physical risk on the job, and workplace culture.

A group-sector composition is characterized by $(\mu_{1w},\mu_{1m},\mu_{2w},\mu_{2m})$,
where~$\mu_{jt}$ is the mass of group~$t$ individuals in sector~$j$, and~$\mu_t$ is the total mass of group~$t$ individuals. Since
$(\mu_{1w},\mu_{1m},\mu_{2w},\mu_{2m})=(\mu_{w}r_w,\mu_{m}r_m,\mu_{w}(1-r_w),\mu_{m}(1-r_m)),$ with~$(r_w,r_m):=(\mu_{1w}/\mu_w,\mu_{1m}/\mu_m)$,
we see that~$(r_w,r_m)$ characterizes a group-sector composition. In what follows, we will therefore refer to~$(r_w,r_m)$ as a composition. 

\begin{definition}
\label{def:eff}
For each group~$t\in\{w,m\}$, call~$r_t:=\mu_{1t}/\mu_t$ the share of people in subpopulation~$t$ who chose sector~$1$, and call~$(r_w,r_m)$ a composition.
\end{definition}

A useful benchmark is the standard Roy model of sector choice, where the non-pecuniary component is absent from the model, i.e., $h_t$ is identically zero. The earnings-only compositions of definition~\ref{def:eo} below are those compositions that would result from maximization of income only, since individuals with~$\Delta\leq0$ (resp.~$\Delta>0$) would choose sector~$1$ (resp. sector~$2$). Without composition preferences, sector group compositions are determined by primitive potential income distributions only.

\begin{definition}
\label{def:eo}
We call earnings-only composition~$(r^e_w,r^e_m)$, where~$r^e_t = \mbox{Pr} \left[ \Delta_t > 0 \right]$, each~$t\in\{w,m\}.$
\end{definition}

To analyze the effects of affirmative action in this model, we choose parametric specifications for the distributions of sector~$1$ advantage~$\Delta_t=Y_{1t}-Y_{2t}$, $t\in\{w,m\}$, and for the composition preference functions~$h_w$ and~$h_m$.

\begin{assumption}
\label{ass:par}
There are parameters~$\rho_t\in(0,1)$, $(c_t,C_t)\in\mathbb R_+^2$ and~$\beta>0$ such that the composition preference function~$h_t$ and the cumulative distribution function~$F_{\Delta_t}$ of group~$t$ sector~$1$ advantage~$\Delta_t=Y_{1t}-Y_{2t}$ satisfy, for each~$u\in(0,1)$ and for~$t\in\{w,m\}$, 
\begin{eqnarray}
\label{eq:para}
h_t(u)=\frac{c_t}{u} \; \mbox{ and } \; F_{\Delta_t}^{-1}(1-u) = C_t\frac{\rho_t-u}{[u(1-u)]^\beta}.
\end{eqnarray}
\end{assumption}
The preference for composition is hyperbolic, which is intended to model strong aversion to being in a small minority. The aversion increases with larger values of the parameter~$c_t$. 
The location of the distribution of sector~$1$ advantage~$\Delta_t=Y_{1t}-Y_{2t}$ is governed by~$\rho_t$. The latter is equal to the earnings-only composition~$r_t^e$ of definition~\ref{def:eo}, i.e., the proportion of group~$t$ individuals choosing sector~$1$ in a Roy model without composition effects. In what follows, we shall use~$r_t^e$ to denote the parameter~$\rho_t$ to avoid proliferation of notation. 

The scale of the distribution of sector~$1$ advantage is governed by~$C_t$. Larger values of~$C_t$ correspond to greater heterogeneity in sector~$1$ advantage among group~$t$ individuals. Parameter~$\beta$ governs the thickness of the tails of the distribution of sector~$1$ advantage. If~$\beta>1$, the tails of the distribution of sector~$1$ advantage are fat enough to counteract aversion for being in a small minority. In other words, there are sufficiently many women with sufficiently large draws of sector~1 advantage to counteract aversion to being in a small minority in sector~1. Hence, the system is pushed away from full segregation (women excluded from sector~1). When~$0<\beta<1$ however, the tails of the distribution of sector~$1$ advantage are not fat enough to counteract aversion for being in a small minority, so that full segregation can occur. The case~$\beta=1$ balances the two effects to yield closed form solutions for equilibrium compositions.

The ratio~$c_t/C_t$ measures the relative importance of pecuniary and composition considerations in choice. Equilibrium configurations depend on the relative importance of pecuniary and composition considerations in choice. We use a convenient reparameterization to gauge this relative importance.
\begin{eqnarray}
\label{eq:gamma}
\begin{array}{lll}
\gamma_w:=\frac{\mu_m}{\mu_w}\frac{c_w}{C_w} & \mbox{and} & \gamma_m:=\frac{\mu_w}{\mu_m}\frac{c_m}{C_m}.
\end{array}
\end{eqnarray}
Parameter~$\gamma_w$, for instance, increases with the relative scale~$c_w/C_w$ of composition versus pecuniary incentives for women, and decreases as the relative labor market participation of women increases.


\subsection{Equilibrium and stability}

The system is in equilibrium when the group composition in each sector is compatible with individual choices. We define it formally as a composition such that no small positive mass of individuals has an incentive to switch sectors.
For an interior equilibrium~$(r_w^*,r_m^*)$, i.e., such that~$(r_w^*,r_m^*)\in(0,1)^2$, this is equivalent to the simple consistency requirement\footnote{Interior equilibrium are Cournot-Nash according to the definition of \citet{M-C:84}.}
\[
\mu_{1t}^*/\mu_t  =  \mbox{Pr} \left[ Y_{1t}-h_{t}\left(\mu_{1t}^*/\mu_1^*\right)>Y_{2t}-h_{t}\left(\mu_{2t}^*/\mu_2^*\right) \right], \mbox{ for } t\in\{w,m\},
\]
where~$\mu_{1t}^*=\mu_tr_t^*,$ $\mu_{2t}^*=\mu_t(1-r_t^*)$, $\mu_j^*=\mu_{jw}^*+\mu_{jm}^*$, $j\in\{1,2\}$, and where the probability is computed with respect to the distribution of potential incomes. After some rearrangement, this is equivalent to
\begin{eqnarray}
F_{\Delta_w}^{-1}(1-r_w^*) & = & g_w(r_w^*,r_m^*,1/r),  \label{eq:eq1} \\
F_{\Delta_m}^{-1}(1-r_m^*) & = & g_m(r_m^*,r_w^*,r), \label{eq:eq2}
\end{eqnarray}
where~$\Delta_t:=Y_{1t}-Y_{2t}$, $r_t^*:=\mu_{1t}^*/\mu_t$, $r:=\mu_w/\mu_m$, and~$g_t$ is defined for each~$(x,y,z),$ by
\begin{eqnarray}
\label{eq:g}
g_t(x,y,z):=h_t\left( \frac{1}{1+z\frac{y}{x}}\right)-h_t\left( \frac{1}{1+z\frac{1-y}{1-x}}\right).
\end{eqnarray}

In the case of a corner equilibrium, the requirement is that the tail of the distribution of sector~1 advantage is not fat enough to overwhelm the disutility of being in a very small minority. For example,~$(r_w^*=0,r_m^*>0)$ can only be an equilibrium composition if the tail of group~$w$'s sector~$1$ advantage~$F_{\Delta_w}^{-1}(1)$ is smaller than the disutility~$g_w(0,r_m^*,1/r)$ of being in a small minority.

\begin{definition}[Equilibrium]
\label{def:eq}
An equilibrium in the Roy model with composition preferences is a composition~$(r_w^*,r_m^*)$ with the following properties.
\begin{enumerate}
\item If~$r_w^*$ (resp. $r_m^* $) is in~$(0,1)$, then~(\ref{eq:eq1}) (resp.~(\ref{eq:eq2})) holds.
\item If~$r_w^*=0$, then  $F_{\Delta_w}^{-1}(1)\leq g_w(0,r_m^*,1/r)$.
\item If~$r_m^*=1$, then  $F_{\Delta_m}^{-1}(0)\geq g_m(1,r_w^*,r)$.
\item Symmetric statements hold for the cases~$r_w^*=1$ and~$r_m^*=0$.
\end{enumerate}
\end{definition}

The equilibrium of definition~\ref{def:eq} is in the sense of Cournot-Nash: No positive mass of individuals has an incentive to deviate. In the case of interior equilibria, condition~(1) says that comparative advantage exactly balances composition preferences for all individuals in the labor market. In the case of corner equilibria, conditions~(2)-(4) say that no positive mass of individuals from  the segregated group has sufficient comparative advantage to switch to the sector, where they are not represented yet.

For example, in the case, where~$F_{\Delta_w}^{-1}(1-r^*)=F_{\Delta_m}^{-1}(1-r^*)=0,$ composition~~$(r^*,r^*)$ is an equilibrium. Indeed, from the definition of~$g_t$, we see that~$g_t(x,y,z)=0$ in that case. In other words, if the median sector~$1$ advantage is the same for both groups, sector group composition may be unaffected by composition preferences. 

We focus on stable
equilibrium, where a deviation from equilibrium choice by a small mass of individuals from one group does not
induce more individuals to deviate. Formally, this corresponds to the following definition\footnote{This form of stability, which is sufficient for our purposes, is weaker than {\em local asymptotic stability} in the theory of dynamical systems, which would require stability in the event of any small joint deviations from both groups of individuals. See for instance the definition page 33 of \cite{BM:89}.}.
\begin{definition}[Stability]
\label{def:stable}
An equilibrium composition~$(r^\ast_w,r^\ast_m)$ is called stable if there is a~$\bar\delta>0$ such that for all~$0<\delta<\bar\delta$, for all $t,t'\in\{w,m\}$ with $t\neq t'$, the following holds:
\begin{eqnarray*}
\begin{array}{lcllcl}
F_{\Delta_t}^{-1}(\max\{0,1-r_t^*-\delta\}) & \leq & g_t(\min\{1,r_t^*+\delta\},r_{t'}^*,r(t)), \\
F_{\Delta_t}^{-1}(\min\{1,1-r_t^*+\delta\}) & \geq & g_t(\max\{0,r_t^*-\delta\},r_{t'}^*,r(t)),
\end{array}
\end{eqnarray*}
where $r(w)=1/r$ and $r(m)=r$ on the right-hand side.
\end{definition}
The force underlying stability of an interior equilibrium can be understood as follows. Take a sufficiently small deviation~$(r_w^\ast-\delta_w,r_m^\ast)$ from equilibrium~$(r_w^\ast,r_m^\ast)$, with~$0<\delta_w<\bar\delta$. 
Suppose this deviation is induced by the small mass of women who are most likely to benefit from switching from sector~1 to sector~2. The latter are therefore ranked between~$1-r_w^\ast$ and~$1-r_w^\ast+\delta$ in the distribution of sector~1 advantage~$\Delta_w$. Hence, their sector~$1$ advantage is between~$F_{\Delta_w}^{-1}(1-r_w^*+\delta) $ and~$F_{\Delta_w}^{-1}(1-r_w^*)$, hence larger than~$g_w(r_w^*-\delta,r_m^*,1/r)$ by definition of stability. Therefore, they are better off returning to sector~1.




\section{Amplification of occupational segregation}


\subsection{Generic existence of an amplified equilibrium}

Composition preferences may induce greater occupational segregation than would result from differences in potential income distributions across groups~$w$ and~$m$.  Once we add composition preferences, occupational segregation may be amplified, in the sense that group imbalances caused by differences in potential incomes may be increased by individual decisions driven by aversion to being in a minority\footnote{Note that results in this section do not rely on the parameterization of assumption~2.}. 

\begin{definition}[Amplified equilibrium]
We call an equilibrium~$(r_w^\ast,r_m^\ast)$ amplified when~$r_w^*<r^e_w<r^e_m<r_m^*$ or $r_m^*<r^e_m<r^e_w<r_w^*$.
\end{definition}

We also uncover a flipping phenomenon, where sufficiently strong composition preferences give rise to a {\em contrarian equilibrium}, where equilibrium compositions reverse pecuniary comparative advantage. Suppose, for instance, that women have a greater sector advantage in finance, but are kept away by a distaste for the socializing practices in that sector (think of visits to strip clubs).
\begin{definition}[Contrarian equilibrium]
\label{def:cont}
A contrarian equilibrium is an equilibrium composition~$(r_w^\ast,r_m^\ast)$ such that,~$\left( r_m^\ast-r_w^\ast \right) \left( r_m^e-r_w^e \right)<0$.
\end{definition}
In a contrarian equilibrium, women are under-represented in sector~1 ($r_w^\ast<r_m^\ast$) in spite of their relative advantage ($r_w^e>r_m^e$), or vice-versa. The intuition is one of coordination. If too few women are in sector~1, women in sector~2 don't find it beneficial to switch to sector~1 in spite of their relative advantage, as composition preferences outweigh pecuniary incentives.

The following proposition shows that the amplification phenomenon is generic under the following very mild additional regularity assumption.

\begin{assumption}\label{ass:reg2}
For any $\bar{r}_m$ and~$\bar r_w\in[0,1]$, the functions
$r_w\mapsto F^{-1}_{\Delta_w}(1-r_w)- g_w(r_w,\bar{r}_m,1/r)$ and $r_m\mapsto F^{-1}_{\Delta_m}(1-r_m)- g_m(r_m,\bar{r}_w,r)$ change sign at most a finite number of times on~$(0,1)$.
\end{assumption}

Under assumption~\ref{ass:reg2}, we show that the amplification phenomenon is generic in the sense that a stable amplified equilibrium always exists, and any non amplified equilibrium is contrarian if the two groups have different sector~1 advantage, i.e.,~$r_w^e\ne r_m^e$. We also give conditions for uniqueness, when the non pecuniary incentives are weak enough.

\begin{proposition}[Amplification effect]
\label{prop:ampli}
Suppose assumption~\ref{ass:reg} holds. Assume that the earnings-only compositions~$r^e_w\ne r^e_m$ are in~$(0,1)$. Then the following hold:

\noindent\underline{Existence:} An equilibrium is either amplified or contrarian. There always exists an amplified equilibrium. Moreover, if assumption~\ref{ass:reg2} holds, there always exists a stable amplified equilibrium. 

\noindent\underline{Uniqueness:} Suppose~$h_t$ is  parameterized as $h_t=c_t\tilde{h}_t$, where $\tilde{h}_t$ is a decreasing function and~$c_t$ is a constant. Then for any~$\bar\varepsilon\in(0,1/2)$, there is an $\varepsilon<\bar\varepsilon$  and a pair~$(\bar c_w,\bar c_m)$ such that for~$c_w\in(0,\bar c_w)$ and~$c_m\in(0,\bar c_m)$, there exists a unique equilibrium with compositions in~$[\varepsilon,1-\varepsilon]^2$. This equilibrium is amplified and stable.
\end{proposition}

\begin{figure}
\caption{Illustration of the amplification of occupational segregation. On the $x$-axis are the compositions~$u$ from~$0$ to~$1$. On the $y$-axis are the values of the quantile functions (for both groups) and the~$g$ preference functions defined in~(\ref{eq:g}) (for both groups), as functions of their first argument, with second argument fixed at the equilibrium value.}
\vskip20pt
\label{fig:ampli}
\includegraphics[scale=0.5]{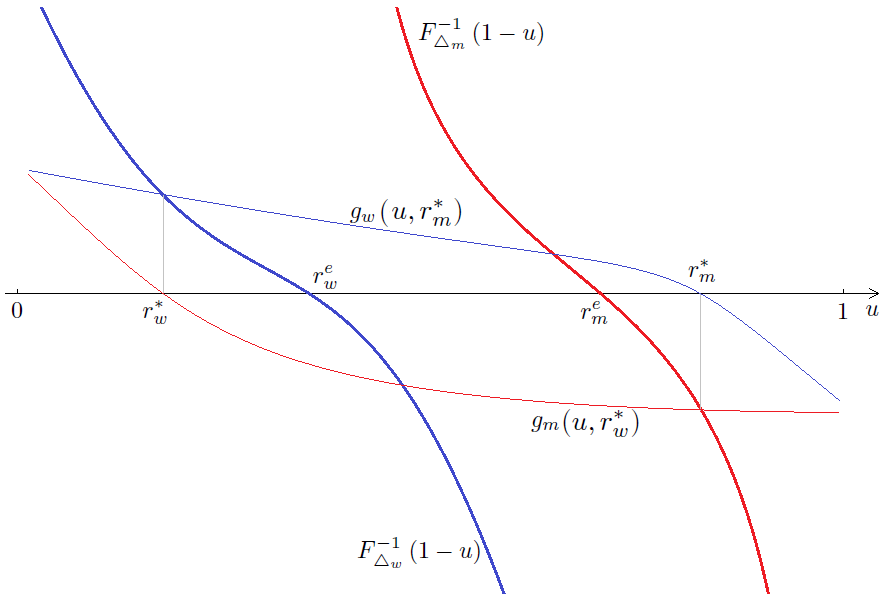}
\end{figure}

Figure~\ref{fig:ampli} gives a graphical illustration of the amplification effect of proposition~\ref{prop:ampli}. The bright blue (resp. red) line traces the quantile function of group~$w$ (resp. group~$m$) sector~$1$ advantage\footnote{The quantiles and~$g$ functions in figure~\ref{fig:ampli} are not derived from a specific parametric specification, but are hypothetical curves drawn for illustrative purposes.} as a function of~$u\in(0,1)$. 
The light blue (resp. red) line traces function~$g_w$ (resp.~$g_m$) as a function of its first argument, with second argument fixed at the equilibrium value~$r_m^\ast$ (resp.~$r_w^\ast$). When $g_w=g_m=0$, the only equilibrium is the earnings-only compositions $(r_w^e,r_m^e)$, from definition~\ref{def:eo}, which are the median compositions, i.e., where quantile functions cross the $x$-axis. When $g_w, g_m>0$, earnings-only compositions are replaced by more uneven equilibrium compositions~$(r^*_w,r^*_m)$ as individuals respond to composition preferences. For instance,~$g_w(u,r_m^*)=g_w(u,r_m^*,1/r)$ is the non-pecuniary gain of a group~$w$ individual from switching to sector~$2$ when the fraction of group~$w$ individuals who choose sector~$1$ is~$u$, and the fraction of group~$m$ individuals who choose sector~$1$ is~$r_m^*$. Equilibrium composition~$r_m^\ast$ is the composition~$u$ such that~$F^{-1}_{\Delta_m}(1-u)=g_m(u,r_w^\ast)$ by equilibrium condition~(\ref{eq:eq2}). Note that it is not coincidental that~$g_w(u,r_w^\ast)$ crosses the $x$-axis at that same value~$u=r_m^\ast$, because~$g_w(u,u)=0$ by construction. Symmetric statements hold for composition~$r_w^\ast$.


\subsection{Amplification with perfectly symmetric groups}

Composition preferences compete with pecuniary incentives in a way that can lead to uneven compositions even when both groups have identical primitive potential income distributions and identical preferences. Suppose both groups have identical mass~$\mu_w=\mu_m$ and identical sector~$1$ advantage, i.e.,~$F_{\Delta_w}=F_{\Delta_m}:=F_\Delta$. Hence, earnings-only compositions are also equal, i.e.,~$r^e_w=r^e_m:=r^e$. Suppose also that both groups also have identical composition preferences.  

In such a configuration, composition~$(r^e,r^e)$ is an equilibrium. However, it is not stable if the slope of~$g$ is steeper than the slope of~$F_{\Delta}^{-1}$ at that point. Indeed, a small mass of group~$w$ individuals switching from sector~$1$ to sector~$2$ would result in incentives for more group~$w$ individuals to switch to sector~$2$ and for group~$m$ individuals to switch to sector~$1$.
This is formalized in the following proposition, the proof of which follows from definitions~\ref{def:eq} and~\ref{def:stable}.

\begin{proposition}\label{prop:sym}
Suppose both groups have identical mass~$\mu_w=\mu_m$, identical sector advantage~$F_{\Delta_w}=F_{\Delta_m}$ and identical composition preferences~$h_w=h_m$. Then the following hold.
\begin{enumerate}
\item Balanced composition~$(r_w^\ast,r_m^\ast):=(r^e,r^e)$ is an equilibrium. This equilibrium is (un)stable if~
$-1/F^\prime_{\Delta_w}(0)< (>)\, h_w^\prime(0.5)/4r^e(1-r^e)$.
\item Uneven composition~$(r_w^\ast,r_m^\ast)$ with~$0<r_w^\ast<r_m^\ast<1$ is an equilibrium if and only if~$F^{-1}_{\Delta_w}(1-r_w^\ast)=g_w(r_w^\ast,r_m^\ast,1)$ and $F^{-1}_{\Delta_m}(1-r_m^\ast)=g_m(r_m^\ast,r_w^\ast,1)$. This equilibrium is stable if~$-1/F^\prime_{\Delta_w}(F^{-1}_{\Delta_w}(1-r_w^\ast)) < \partial g_w/\partial r_w \, (r_w^\ast,r_m^\ast,1)$ and $-1/F^\prime_{\Delta_m}(F^{-1}_{\Delta_m}(1-r_m^\ast)) < \partial g_m/\partial r_m \, (r_m^\ast,r_w^\ast,1)$.
\end{enumerate}
\end{proposition}

\begin{figure}
\caption{Illustration of purely endogenous composition imbalances. On the $x$-axis are the compositions~$u$ from~$0$ to~$1$. On the $y$-axis are the values of the quantile functions (identical for both groups) and the~$g$ preference functions defined in~(\ref{eq:g}) (identical for both groups), as functions of their first argument, with second argument fixed at the equilibrium value.}
\vskip20pt
\label{fig:multiple}
\includegraphics[scale=0.55]{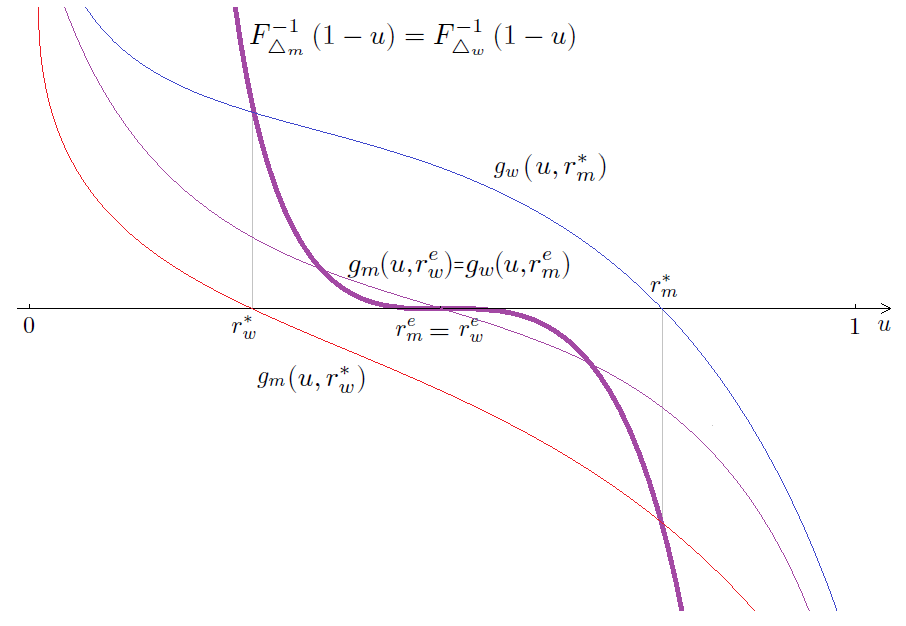}
\end{figure}

Figure~\ref{fig:multiple} illustrates this situation. The figure is identical to figure~\ref{fig:ampli} except that the quantile functions of both groups are identical, hence~$r_m^e=r_w^e$, and identical composition preference function~$g_m(u,r_w^e)=g_w(u,r_m^e)$ are also traced on the figure. The slope of the composition preference effect~$g$ is steeper at fraction~$r^e:=r_m^e=r_w^e$ than the slope of the quantile function~$F^{-1}_{\Delta_t}$. The parity equilibrium composition~$(r^e,r^e)$ is unstable, whereas the equilibrium~$(r_w^*,r_m^*)$ with unbalanced group compositions is stable. 
Despite both groups being identical in preference and potential incomes, and despite no sector being favored a priori, composition preferences push the population to uneven compositions in both sectors. 


\subsection{Amplified and contrarian equilibrium}

We impose the parametric specification of assumption~\ref{ass:par}, so that we can quantify the amplification effect and identify conditions required on the intensity of pecuniary and composition motives for an interior amplified equilibrium to exist and to be unique. 

We start with the case $\beta=1$, which provides closed-form solutions and the ability to enumerate all equilibria. We then extend the analysis to cases $\beta>1$ and $\beta<1$, where we characterize when there is a unique stable amplified equilibrium and when a contrarian equilibrium exists.

\subsubsection{Case~$\beta=1$}

\begin{figure}
\caption{Characterization of all equilibria under assumption~\ref{ass:par} with~$\beta=1$ and~$r_w^e<0.5<r_m^e$. In each region, the equilibria are given in the form~$(r_w^*,r_m^*)$.} 
\vskip30pt
\label{fig:eq-reg}
\includegraphics[scale=0.34]{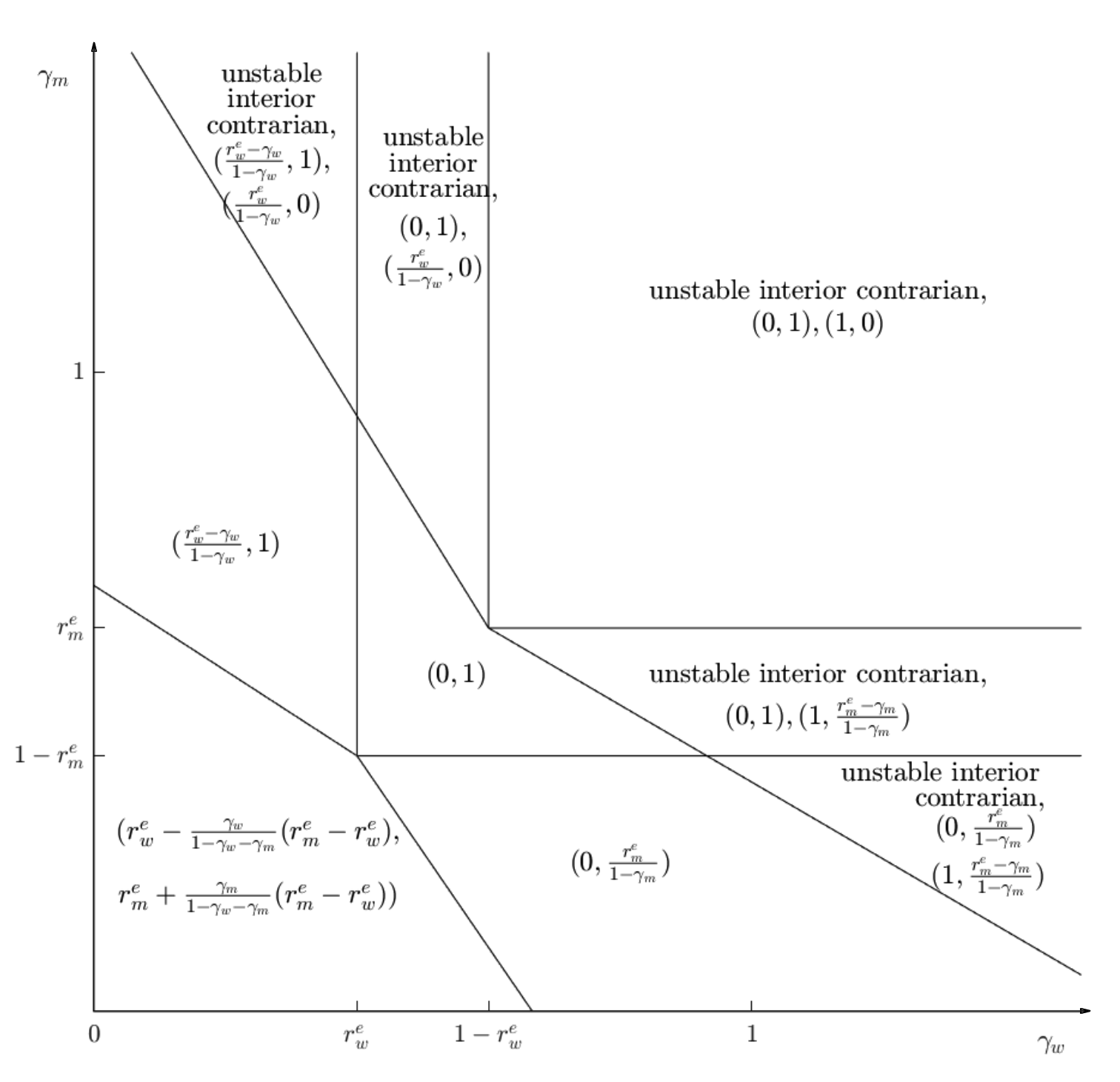}
\end{figure}

In case~$\beta=1$, the thickness of the tail of the distribution of sector~$1$ advantage~$F_{\Delta_t}$, which regulates the amount of exceptionally talented individuals, is perfectly balanced by the intensity of the dislike of being in a small minority. In this case, we obtain uniqueness of stable equilibrium, which is amplified, and which is given in closed form. Equilibrium regions are characterized graphically in figure~\ref{fig:eq-reg} in the case~$r_w^e<0.5<r_m^e$. Other cases produce similar equilibrium profiles. The characterization of all equilibria in case~$\beta=1$ reveals a large region, where a stable amplified equilibrium exists and is unique, as formalized in the following proposition.

\begin{proposition}
\label{prop:1}
Suppose assumption~\ref{ass:par} holds with~$\beta=1$. Without loss of generality, let~$r_w^e \leq r_m^e$. 
(1)~If~$\gamma_w+\gamma_m<1$, there is a unique equilibrium, and it is stable and amplified. (2)~If, in addition, $\bar\gamma:=\max\{\gamma_wr_m^e/r_w^e+\gamma_m,\gamma_w+\gamma_m(1-r_w^e)/(1-r_m^e)\}<1$, then the unique stable and amplified equilibrium has compositions given by:
\begin{eqnarray*}
(r_w^*,r_m^*) & = & \left(r_w^e - \frac{\gamma_w}{1-\gamma_w-\gamma_m} ( r_m^e - r_w^e ) \; , \; r_m^e + \frac{\gamma_m}{1-\gamma_w-\gamma_m} ( r_m^e - r_w^e ) \right)\!\!.\hskip25pt
\end{eqnarray*}
\end{proposition}

The amplification effect of proposition~\ref{prop:ampli} is quantified here, given the closed form solution\footnote{The amplification effect we quantify here is the amplification that arises in a stable interior equilibrium. A more extreme form of amplification occurs in corner equilibria, where at least one group is entirely confined to the sector in which it holds a relative advantage.}. Occupational segregation increases with the strength~$\gamma_t$ of both groups' composition preferences relative to pecuniary incentives. Occupational segregation also increases with the primitive mean difference~$r_m^e-r_w^e$ between group~$m$ and group~$w$ sector~$1$ advantage.

Figure~\ref{fig:eq-reg} also shows that if composition preferences are strong, as measured by~$\gamma_w$ and~$\gamma_m$, then there are two stable equilibria, one amplified and the other contrarian and they either involve full segregation or complete segregation. 

\subsubsection{Case~$\beta<1$}

Under assumption~\ref{ass:par} with tail parameter~$\beta<1$, sector~1 advantage is not large enough to overwhelm the disutility of being in a very small minority, unless composition preferences are very weak. Hence multiple corner equilibria arise, where either group is confined to a single sector, or there is complete segregation, where both sectors have a single group of worker. We give a full characterization of the set of equilibria for small enough~$\gamma_m$ and~$\gamma_w$. The set of equilibria is complicated, but the important feature is that away from the boundaries corresponding to full segregation, the only attractor of the dynamic system is an interior stable amplified equilibrium.

\begin{table}
\begin{center}
\begin{tabular}{l|c|c|c} 
   & $0\leq r_w<\varepsilon$ & $\varepsilon\leq r_w\leq 1-\varepsilon$ & $1-\varepsilon<r_w\leq 1$\\
 \hline
 $1-\varepsilon<r_m\leq 1$ & $(0,1)$ & $(r_w^{(1)},1)$ & \\ 
 \hline
 $\varepsilon\leq r_m\leq 1-\varepsilon$ & $(0,r_m^{(3)})$ & $(r_w^{(2)},r_m^{(2)})$ & $(1,r_m^{(1)})$ \\ 
 \hline
$0\leq r_m<\varepsilon$ & & $(r_w^{(3)},0)$ & $(1,0)$
\end{tabular}
\end{center}
\caption{Complete set of stable equilibria in case~$0<\beta<1$.}
\label{tab:17}
\end{table}

\begin{proposition}
\label{prop:beta1}
Suppose assumption~\ref{ass:par} holds with~$\beta\in(0,1)$. Then, complete segregation, where both sectors have a single group of workers, is always a stable equilibrium. In addition, assume, without loss of generality, that the earnings-only compositions satisfy~$r^e_w<r^e_m$, then for any~$\bar\varepsilon\in(0,1/2)$, there is an $\varepsilon<\bar\varepsilon$ and a pair~$(\bar \gamma_w,\bar \gamma_m)$ such that for~$\gamma_w\in(0,\bar \gamma_w)$ and~$\gamma_m\in(0,\bar \gamma_m)$, there are exactly 7 stable equilibria, exactly one of which is interior. The 7 equilibria are given in table~\ref{tab:17}, where~$\varepsilon<r_w^{(1)}<r_w^{(2)}<r_w^e<r_w^{(3)}<1-\varepsilon$ and~$\varepsilon<r_m^{(1)}<r_m^e<r_m^{(2)}<r_m^{(3)}<1-\varepsilon$.
\end{proposition}

The stable amplified central equilibrium, i.e., equilibrium~$(r_w^{(2)},r_m^{(2)})$ in table~\ref{tab:17}, is the unique equilibrium with compositions within the box~$[\varepsilon,1-\varepsilon]^2$ already identified in all generality in proposition~\ref{prop:ampli}. It is also the only stable equilibrium interior equilibrium. The other six stable equilibria are corner equilibria. Two of them,~$(0,1)$ and~$(1,0)$, exhibit complete segregation, i.e., both groups are confined to their own sector. The other four exhibit full segregation of one of the two groups: In equilibria~$(0,r_m^{(3)})$ and~$(1,r_m^{(1)})$, women are all in the same sector, and in equilibria~$(r_w^{(3)},0)$ and~$(r_w^{(1)},1)$, men are all in the same sector.

\subsubsection{Case~$\beta>1$}

Under assumption~\ref{ass:par} with tail parameter~$\beta>1$, there are always enough women whose sector~1 advantage is so large that it overwhelms the disutility of being in a very small minority. Hence, full segregation of a group in a single sector is no longer an equilibrium. In a {\em symmetric environment}, i.e.,  with~$\gamma_w=\gamma_m$ and~$r_w^e=1-r_m^e$, we show that the only equilibria that can survive are one stable amplified equilibrium, and two contrarian equilibria. We also give more general conditions (beyond the symmetric case) under which the stable amplified equilibrium is the unique equilibrium.

\begin{proposition}
\label{prop:1beta}
Suppose assumption~\ref{ass:par} holds with~$\beta>1$, and the earning-only compositions satisfy~$0<r_w^e<r_m^e<1$. Then, there are no corner equilibria. Moreover, there is a pair~$(\bar \gamma_w,\bar \gamma_m)$ such that for~$\gamma_w\in(0,\bar \gamma_w)$ and~$\gamma_m\in(0,\bar \gamma_m)$, there is a stable amplified equilibrium and it is the only equilibrium. Finally, in the symmetric environment, i.e., when~$\gamma_w=\gamma_m$ and~$r_w^e=1-r_m^e$, for each~$\gamma_w>1-r_w^e$, there is a~$\bar\beta(\gamma_w)>1$ such that
\begin{enumerate}
\item If~$\gamma_w<1-r_w^e$ or~$\beta>\bar\beta(\gamma_w)$, there is a unique equilibrium, and it is stable and amplified. 
\item If~$\gamma_w>1-r_w^e$ and ~$\beta<\bar\beta(\gamma_w)$, then there are exactly three equilibria. One stable amplified equilibrium, and two contrarian equilibria. 
\end{enumerate}
\end{proposition}

Proposition~\ref{prop:1beta} shows that a high incidence of individuals with large sector advantage rules out corner equilibria. The pecuniary incentive is strong enough to pull a large enough minority away from full segregation. A strong enough pecuniary incentive relative to composition preferences also rules out the contrarian equilibria, so that there exists a unique equilibrium, which is stable and amplified. Finally, proposition~\ref{prop:1beta} also gives a full characterization of equilibria in the symmetric environment. This characterization is illustrated in case~$r_w^e=1-r_m^e=0.4$ on figure~\ref{fig:cata} below.

\begin{figure}
\caption{Equilibrium regions in~$\gamma_w-\beta$ space in case~$\beta>1$, $\gamma_w=\gamma_m$ and~$r_w^e=1-r_m^e=0.4$.}
\vskip30pt
\label{fig:cata}
\includegraphics[scale=0.20]{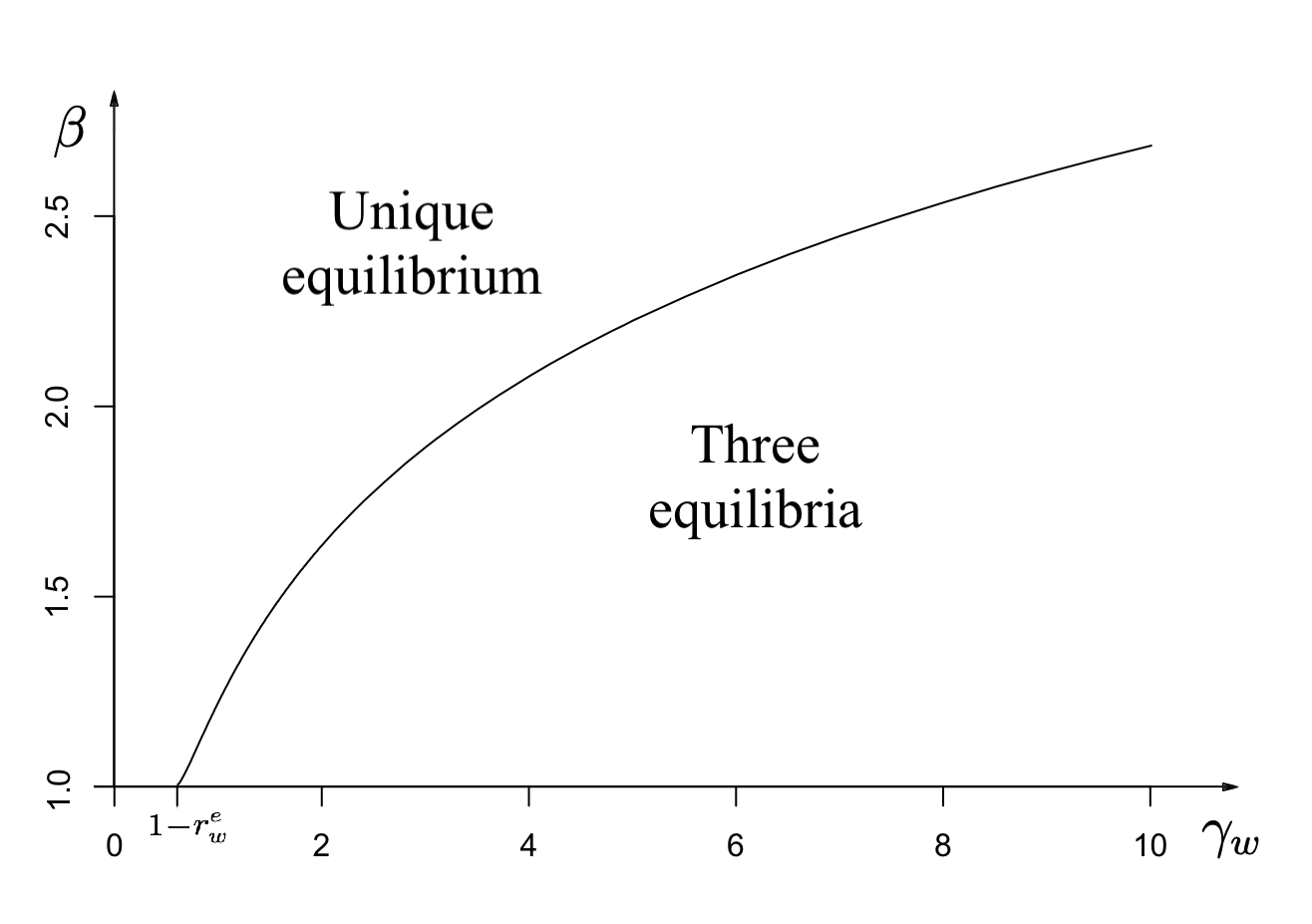}
\end{figure}




\section{Tipping and affirmative action}
\label{sec:tip}

The Roy model with composition preferences displays tipping phenomena of two kinds. One type of tipping effect occurs when a small deviation in~$(r_w,r_m)$ from one equilibrium tips the system to a different equilibrium. The second type of tipping effect occurs when the set of equilibria changes abruptly as the parameters~$(\beta,\gamma_w,\gamma_m)$ that govern the relative strength of pecuniary and non-pecuniary incentives cross a threshold. Unlike residential tipping in the work of \citeauthor{Schelling:71} (\citeyear{Schelling:71,Schelling:73,Schelling:78}), tipping in this model tends to be beneficial and to occur from highly segregated states to more even ones, or from contrarian states (where compositions reverse relative advantage) to non contrarian ones. Each type of tipping in this model is associated with a type of policy. Tipping as a shift between equilibrium, which we call {\em nudge tipping}, is triggered by policies operating directly on compositions, such as quotas. Tipping as a change in the set of equilibria, which we call {\em catastrophic tipping}, is triggered by policies operating on preferences or incomes, such as subsidies or changes in amenities, or changes in labor force participation. In this section, we define the two kinds of tipping and draw implications about each type of tipping from propositions~\ref{prop:beta1} and~\ref{prop:1beta}.


\subsection{Nudge tipping}
\label{sec:quotas}

The first kind of tipping we consider is a jump from one equilibrium to another following a forced shift of the compositions from the domain of attraction of the origin equilibrium to the domain of attraction of the destination equilibrium, as formalized below. We define this kind of tipping formally as follows.

\begin{definition}[Nudge tipping]
\label{def:nudge}
We call {\em nudge tipping} a shift from a stable corner equilibrium to a stable interior equilibrium.
\end{definition}

\begin{figure}
\caption{Phase diagram for dynamical system~(\ref{eq:phase}) with~$\beta=0.5, \gamma_w=\gamma_m=0.1$, $r^e_w=0.4$ and~$r^e_m=0.6$. The curves are the loci, where $\dot{r}_w=\dot{r}_m=0$ in equations~(\ref{eq:phase}). Stable equilibria are marked by red dots. There are~$6$ stable corner equilibria, and~$1$ stable interior (mildly) amplifying equilibrium. Black arrows indicate the direction of attraction of the differential system. Tipping occurs when a small deviation from any of the stable corner equilibria pushes the system to the domain of attraction of the stable interior amplifying equilibrium.} 
\vskip20pt
\label{fig:phase}
\includegraphics[scale=0.45]{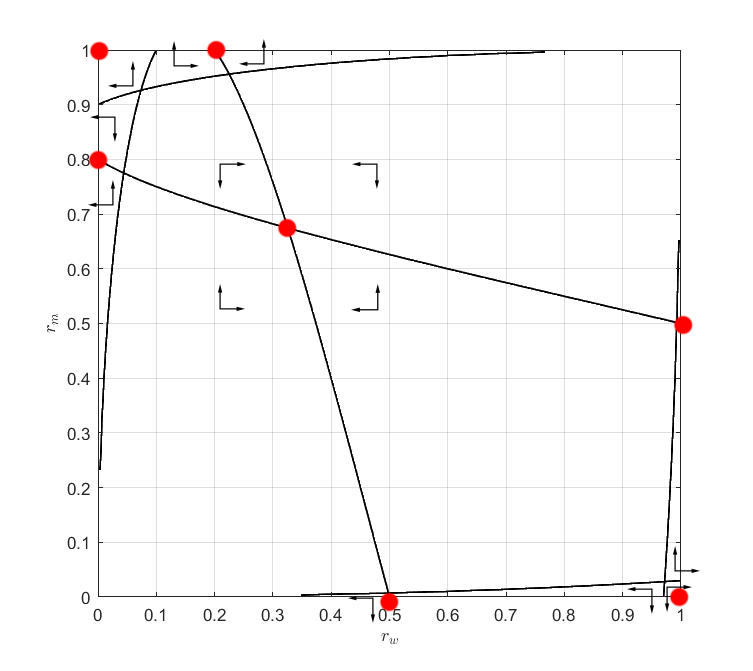}
\end{figure}

To analyze tipping between two equilibria, we need to introduce the notion of domain of attraction of an equilibrium. Consider the dynamical system induced by the Roy model with composition preferences. Take group~$w$ individuals for instance. If
\[
\mu_{1w}/\mu_w  >  \mbox{Pr} \left[ Y_{1w}-h_{w}\left(\mu_{1w}/\mu_1\right)>Y_{2w}-h_{w}\left(\mu_{2w}/\mu_2\right) \right],
\]
or, equivalently, if~$F^{-1}_{\Delta_w}(1-r_w) <  g_w(r_w,r_m,r)$,
the system is not in equilibrium, and a part of the group~$w$ population is induced to move from sector~$1$ to sector~$2$.
This also affects group~$m$ individuals, so that movements of both groups of individuals have to be considered jointly. Formally, the dynamical system is given by
\begin{eqnarray}
\label{eq:phase}
\begin{array}{lll}
\dot{r}_w & = & F^{-1}_{\Delta_w}(1-r_w) -  g_w(r_w,r_m,r),  \\ \\
\dot{r}_m & = & F^{-1}_{\Delta_m}(1-r_m) - g_m(r_m,r_w,1/r), 
\end{array}
\end{eqnarray}
where~$\dot{r}$ indicates the time derivative of~$r$. Figure~\ref{fig:phase} is a phase diagram for this dynamical system in the case of~$\beta=0.5$ in assumption~\ref{ass:par} and specific values for the earnings-only compositions~$r^e_w$ and~$r^e_m$ and the relative strength~$(\gamma_w,\gamma_m)$ of pecuniary and non-pecuniary incentives. Blue curves are the loci of~$F^{-1}_{\Delta_w}(1-r_w) =  g_w(r_w,r_m,r)$ and~$F^{-1}_{\Delta_m}(1-r_m) = g_w(r_m,r_w,1/r)$. Black arrows indicate the direction of change in~$(r_w,r_m)$ off equilibrium. Red dots indicate stable equilibria. 

Figure~\ref{fig:phase} illustrates a configuration where nudge tipping can happen. Since all but extreme compositions lie within the domain of attraction of the interior (mildly) amplified equilibrium, a very low quota of women in the sector, where they are not represented, pushes the composition into the domain of attraction of the interior equilibrium. The system dynamics then pushes compositions to the interior equilibrium as a result of individual incentives only. Therefore, a very small quota of women in a sector induces nudge tipping as in definition~\ref{def:nudge}. Proposition~\ref{prop:nudge} below (which is a corollary of lemma~\ref{lem:17*} in the appendix), gives some conditions under which nudge tipping occurs.

\begin{proposition}
\label{prop:nudge}
Suppose assumption~\ref{ass:par} holds with~$\beta\in(0,1)$, $\gamma_w=\gamma_m<\bar\gamma$ and~$r_w^e=1-r_m^e$, where $\bar\gamma$ is defined in proposition~\ref{prop:beta1}. There exists a minimal~$\underbar q\in(0,0.5)$ such that any quota~$q\in(\underbar q,0.5)$ of each group in each sector triggers nudge tipping. The minimal quota~$\underbar q$ is increasing in~$\gamma_w=\gamma_m$ and decreasing in~$\beta$.
\end{proposition}

Nudge tipping doesn't involve shifts in parameters, or variation in the fundamentals of the model. As illustrated in figure~\ref{fig:phase}, holding parameters~$\beta$, $(r^e_w,r^e_m)$ and~$(\gamma_w,\gamma_m)$ fixed, hence holding both sector~1 advantage and preferences fixed for both groups, a very small quota for the representation of each group in each sector pushes the dynamical system from any of the corner equilibria to the unique interior equilibrium~$(r^*_w,r^*_m)$. The latter displays the amplification feature uncovered in proposition~\ref{prop:ampli}, i.e.,~$r^*_w<r^e_w<r^e_m<r^*_m$. However, given the modest composition preferences relative to pecuniary incentives (low values of parameters~$\gamma_w$ and~$\gamma_m$), this amplification effect is small.

Although nudge tipping, as noted above, doesn't involve shifts in parameters, the latter can affect the magnitude of the quotas necessary to trigger it. A larger value of~$\beta$, hence more highly talented individuals, strengthens the pecuniary incentives in sector choice. A smaller value of~$\gamma_w$, hence a weaker dislike of being in a small minority, weakens the non pecuniary incentives in sector choice. Both, therefore, enlarge the domain of attraction of the interior equilibrium and reduce the magnitude of the necessary quota to trigger tipping.


\subsection{Catastrophic tipping}
\label{sec:subsidies}

The second kind of tipping phenomenon we consider is induced by the disappearance of a stable equilibrium. This kind of tipping is associated with the notion of tipping point in catastrophe theory. Tipping in \citeauthor{Schelling:71} [\citeyear{Schelling:69,Schelling:71,Schelling:73,Schelling:78}], \citet{CMR:2008} and \citet{Pan:2015} falls in that category.

\begin{definition}[Catastrophic tipping]
\label{def:cata}
We call {\em catastrophic tipping} the disappearance of a stable equilibrium as a result of a parameter shift. \end{definition}

Unlike nudge tipping, catastrophic tipping involves no exogenous shift in compositions. There is a point of discontinuity in the parameter space, where the set of equilibria changes abruptly. Tipping occurs if this change involves the disappearance of at least one stable equilibrium. In \citet{CMR:2008} and \citet{Pan:2015}, tipping occurs as a result of the disappearance of an interior equilibrium with mixed residential reals estate or labor market compositions, leaving only a segregated equilibrium. In proposition~\ref{prop:cata} (which is a corollary of proposition~\ref{prop:1beta}), we identify conditions under which catastrophic tipping occurs as a result of the disappearance of a contrarian equilibrium, leaving only the stable (non contrarian) amplified equilibrium.

\begin{proposition}
\label{prop:cata}
Suppose assumption~\ref{ass:par} holds with~$\gamma_w=\gamma_m$ and~$r_w^e=1-r_m^e$. If~$\gamma_w+r_w^e>1$ and~$\beta<\bar\beta(\gamma_w)$, where~$\bar\beta(\gamma_w)>1$ is defined in proposition~\ref{prop:1beta}, then~$\beta$ crossing the~$\bar{ \beta}(\gamma_w)$ curve in figure~\ref{fig:cata} from below or~$\gamma_w$ crossing the~$\bar{ \beta}(\gamma_w)$ curve from the right both induce catastrophic tipping.
\end{proposition}

\begin{figure}
\caption{Phase diagrams for dynamical system (\ref{eq:phase}) with $\mu_w=\mu_m=1, c_m=c_w=2, C_{w}=C_{m}=1, r_{w}^{e}=0.6$ and $r_{m}^{e}=0.4$. The two diagrams are differ in $\beta$. On the left-hand diagram $\beta=1.5$, on the right-hand diagram $\beta=2$. The curves are the loci, where $\dot{r}_w=\dot{r}_m=0$ in equations~(\ref{eq:phase}).} 
\vskip20pt
\label{fig:noncorn}
\includegraphics[scale=0.4]{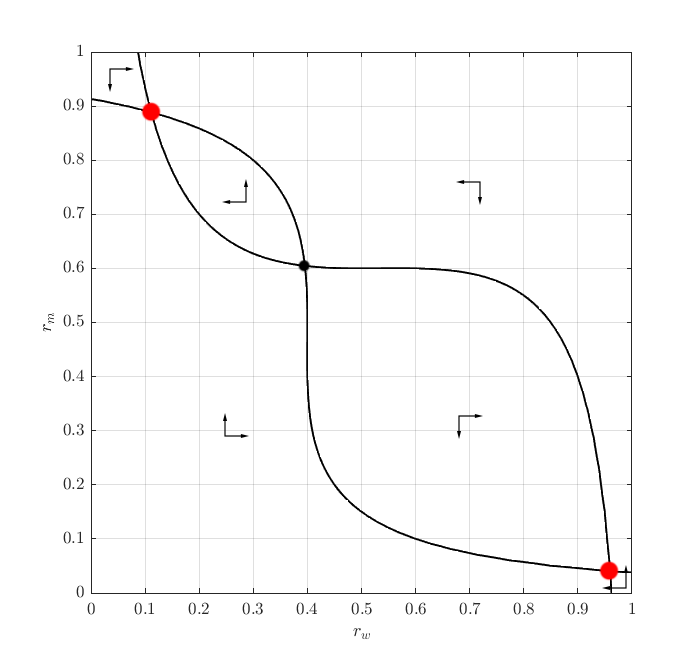}
\includegraphics[scale=0.4]{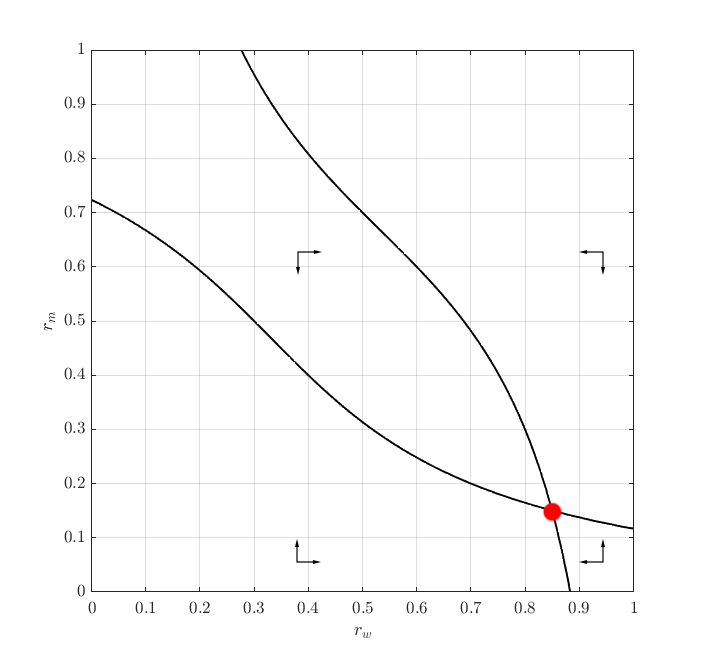}
\end{figure}

Figure~\ref{fig:noncorn} shows an example of disappearance of contrarian equilibria as a result of an increase in~$\beta$. As~$\beta$ increases through a tipping point, the stable contrarian equilibrium disappears and the system tips to the only remaining equilibrium, which is an amplifying interior equilibrium.

The quota policy entertained in section~\ref{sec:quotas} operates through a change in compositions~$(r_w,r_m)$ while keeping the distribution~$F_{\Delta}$ of sector~$1$ advantage and preferences~$h_t$ fixed,~$t\in\{w,m\}$. In contrast, subsidies operate on~$F_{\Delta}$, and amenity changes operate on composition preferences~$h_t$ without affecting compositions directly. Consider a class of policies\footnote{Policies that change workplace culture, such as sexual harassment rules, are part of that class.} that reduce parameter~$\gamma_w$ or increase~$\beta$, which operate on $h_t$ or $F_{\Delta}$ under assumption 2. By definition of~$\gamma_w$, the former includes (a)~subsidies in the sector where women are under-represented (sector~1), i.e., increases in~$C_w$, (b)~improvements in sector~1 amenities, i.e., decrease in~$c_w$, or (c)~increases in labor force participation, i.e., increases in~$\mu_w$. The latter class of policies includes policies that increase the mass of women with large sector~1 advantage, such as merit based scholarships for women in STEM at every stage of the human capital investment. Such policies can provoke tipping through the disappearance of contrarian equilibria. 

If~$r_w^e>r_m^e$ and initial female labor force participation is low, contrarian equilibria can also disappear as female labor force participation increases. The only remaining equilibrium is an interior equilibrium that reflects women's advantage in sector~$1$, i.e., such that~$r_{w}^{*}>r_{m}^{*}$. This provides an alternative mechanism for the rapid shifts in gender compositions for occupations such as bankers and insurance agents documented in \citet{Pan:2015}. 




\section{Empirical content of the Roy model with composition effects}
\label{sec:empirical}

\subsection{Empirical model}

We propose the following empirical version of the Roy model with composition preferences. Individual~$i$ with group~$t_i\in\{w,m\}$ chooses sector~$1$, denoted by~$D_i=1$ if~$\Delta_i>g_{t_i}+\xi_i$, where~$\Delta_i=Y_{1i}-Y_{2i}$, $g_{t_i}:=g_w(r_w,r_m,1/r)$ if~$t_i=w$, and~$g_{t_i}:=g_m(r_m,r_w,r)$ if~$t_i=m$, and~$\xi_i$ is a random variable with cumulative distribution function~$F_\xi$, independent of~$\Delta_i$. Individual~$i$ chooses sector~$2$ otherwise, denoted by~$D_i=2$. We interpret this choice model as the result of utility maximization as in section~\ref{sec:frame}, where the utility includes an additive sector specific random utility term. Finally let~$\underline w\geq0$ be the minimum wage and assume that the support of all income distributions is bounded below by~$\underline w$.

Individual~$i$'s realized income is~$Y_i=Y_{1i}$ when~$D_i=1$ and~$Y_{2i}$ when~$D_i=2$. We assume that the four distributions of realized incomes for each group and each sector are observed in the population. Hence, the quantities~$\mbox{Pr}(Y_i\leq y, D_i = d,t_i=t)$ are assumed known for all~$t\in\{w,m\}$,~$d\in\{0,1\}$, and~$y\geq\underline w$. Group-sector compositions~$(r^*_w,r^*_m)$ are observed. They are assumed to be equilibrium compositions. The ratio~$r$ of total populations of both groups is also observed. The cumulative distribution function~$F_\xi$ of the random utility term is assumed known.  The minimum wage~$\underline w$ is also assumed known.

\subsection{Sharp bounds on the primitives}
\label{sec:sharp}

The distribution of potential incomes is not observed and cannot be directly estimated, since income is only observed in the chosen sector. Therefore, the vector of parameters is not point identified in general\footnote{See \citet{MHM:2020} for a discussion of partial identification issues in the Roy model.}. We provide a sharp characterization of the identified set, i.e., the set of parameter values that cannot be rejected on the basis of the decision model and the observed choices and realized incomes.

Start with simple implications of the model. Take individual~$i$ with group~$t_i$, potential incomes~$(Y_{1i},Y_{2i})$, sector~$1$ advantage~$\Delta_i$, and realized income~$Y_i$.  Suppose individual~$i$ is in sector~$1$ and has a realized income that is smaller than~$y$, for some~$y\geq\underline w$. Individual~$i$'s sector advantage~$\Delta_i=Y_{1i}-Y_{2i}$ must then be smaller than~$y-\underline w$, since~$Y_{1i}=Y_{i}\leq y$ and~$Y_{2i}\geq\underline w$. Moreover, by the choice model,~$D_i=1$ also implies~$\Delta_i>g_{t_i}^*+\xi_i$, where~$g_{t_i}^*:=g_w(r_w^*,r_m^*,r)$ if~$t_i=w$ and~$g_{t_i}^*:=g_m(r_m^*,r_w^*,1/r)$ if~$t_i=m$. Hence, we obtain, for all~$y\geq\underline w$, the implication
$\mbox{Pr}(g_{t_i}^*+\xi_i<\Delta_i\leq y-\underline w,t_i=t) \geq \mbox{Pr}(Y_i\leq y, D_i = 1,t_i=t)$, which gives a bound on the unknown distribution of individual~$i$'s sector~$1$ advantage~$\Delta_i$. Since the random utility component~$\xi_i$ is independent of~$\Delta_i$, and its distribution is known, the bound can be integrated over~$\xi_i$ to yield
\begin{eqnarray}
\label{eq:ec1}
\begin{array}{lll}
\int\mbox{Pr}(g_{t_i}^*+\xi<\Delta_i\leq y-\underline w,t_i=t)dF_\xi(\xi) & \geq & \mbox{Pr}(Y_i\leq y, D_i = 1,t_i=t).
\end{array}
\end{eqnarray}
The same reasoning applies to the case~$Y_i>y$ and~$D_i=2$ to yield the bound
\begin{eqnarray}
\label{eq:ec2}
\begin{array}{lll}
\int\mbox{Pr}(\underline w-y<\Delta_i\leq g_{t_i}^*+\xi,t_i=t)dF_\xi(\xi) & \geq & \mbox{Pr}(Y_i> y, D_i = 0,t_i=t),  
\end{array}
\end{eqnarray}
for all~$y\geq\underline w$, $t\in\{w,m\}$.

\subsection{Inference on structural parameters and counterfactual analysis}

When potential incomes and composition preferences satisfy assumption~\ref{ass:par},
system~(\ref{eq:ec1})-(\ref{eq:ec2}) defines a continuum of moment inequalities.
For each~$y\geq\underline w$, and each value of the parameter vector~$(r^e_w,r^e_m,c_w,c_m,C_w,C_m)$, the moment inequalities can be evaluated numerically, and confidence regions derived using existing inference methods on moment inequality models (see for instance \citet{CS:2017} for a survey).

Once structural parameters are recovered, we can evaluate the effect of policies described in sections~\ref{sec:quotas} and~\ref{sec:subsidies} in the following way. Given the parameter values and the observed equilibrium compositions, the phase diagram for the dynamical system~(\ref{eq:phase}) determines the equilibrium that would result from the imposition of a quota system, holding structural parameters fixed. In the case of a policy that shifts the parameters~$C_t$ (a subsidy) of~$c_t$ (an exogenous change in workplace culture), the new parameter values determine a new phase diagram for the dynamical system~(\ref{eq:phase}), which also determines the resulting equilibrium.




\section{Discussion}

We proposed a model combining comparative advantage with composition preferences in labor market sector selection. This allowed us to propose one mechanism to explain high levels of occupational segregation as well as the observed counterintuitive negative correlation between inequality and gender occupational segregation.
The model could easily be extended to accommodate preference for parity, where the disutility function increases on either side of parity. This could be interpreted as a tradeoff between homophily and marriage market considerations. Group identity (\citet{AK:2000}, \citet{BKP:2015}, \citet{Bertrand:2011}), operates through human capital accumulation, but also compositions. For instance, a small ratio of women reinforces the gendered perception of an occupation. Mentoring in the career development of women operates through potential incomes, as lifetime earnings are affected, but it could be incorporated in a multi generation version of the model).

However, the sharp separation between comparative advantage, whose effect on self-selection is mediated through potential incomes, and non-pecuniary drivers of selection, mediated through compositions and social effects, brings with it some major limitations. One major limitation of the model is that composition effects on learning are ruled out. Potential incomes (that incorporate human capital accumulation) are unaffected by group-sector compositions. Peer effects are limited to non-pecuniary benefits and cannot affect future productivity. The model also rules out effects of gender compositions on the gender wage gap, through realistic channels such as increased wage discrimination in STEM fields because of low representation of women.
The model is a full information game, so that gender specific bias in wage expectations is ruled out as a factor in under-representation of women in STEM. It is also frictionless, so there is no place for job search strategies, homophily in referral networks, as in \cite{FS:2005}, \citet{Zelzer:2020} and \citet{BL:2020}.




\begin{appendix} 


\section{Proofs of results in the main text}


\subsection{Proof of proposition~\ref{prop:ampli}}


\subsubsection{Preliminaries}

Under assumption~\ref{ass:reg}, the function~$g_t(x,y,z)$ defined in~(\ref{eq:g}) for~$t\in\{w,m\}$ is decreasing in~$x$ and increasing in~$y$ and satisfies~$g_t(x,x,z)=0$ for all~$(x,y,z)\in(0,1)^3$. Also $F^{-1}_{\Delta_t}(1-r_t)$ is decreasing in $r_t$.

Define the set~$S:=S_1\cap S_2 \cap [0,1]^2$, where
\begin{eqnarray*}
S_1 & = & \left\{(r_w,r_m)| F^{-1}_{\Delta_w}(1-r_w)\leq g_w(r_w,r_m,1/r)\right\} \\
S_2 & = & \left\{(r_w,r_m)| F^{-1}_{\Delta_m}(1-r_m)\geq g_m(r_m,r_w,r)\right\}.
\end{eqnarray*}
The functions $F^{-1}_{\Delta_t}$ and $g_t$ are continuous for~$t\in\{w,m\}$, 
so~$S_1,$ $S_2$ and~$S$ are closed. 

It can be easily checked that~$S$ satisfies the following three properties.
\begin{eqnarray}
&& \left\{ (r_w^e,r_w^e),(r_w^e,r_m^e),(r_m^e,r_m^e) \right\}\subset S, \nonumber \\
&& \left\{(r_w,r_m)|r_w\geq r_w^e,r_m\leq r_m^e,r_w\leq r_m  \right\}\subset S, \mbox{ and}\nonumber \\
&& \mbox{If }(a_1,b_1)\in S, (a_2,b_2)\in S \mbox{ and } a_1\leq a_2, b_1\leq b_2, \mbox{ then } (a_1,b_2)\in S.
\label{eq:property3}
\end{eqnarray}


\begin{fact}\label{fact:upperleft1}
Let~$K$ be a connected component of~$S$. There exists~$(r_w^L,r_m^U)\in K$ such that for any~$(r_w,r_m)\in K$, $r_w^L\leq r_w$ and~$r_m\leq r_m^U$. We call~$(r_w^L,r_m^U)$ the {\em upperleft corner} of component~$K$.
\end{fact}

\begin{proof}[Proof of Fact~\ref{fact:upperleft1}]
All the connected components of a closed set in~$\mathbb R^2$ are closed. Moreover,~$S$ is bounded. So~$K$ is closed and bounded, hence compact. Thus the Weierstrass Theorem applies to continuous functions~$(r_w,r_m)\rightarrow r_w$ and~$(r_w,r_m)\rightarrow r_m$. Let the minimum value of the first function be $r_w^L$ and let it be attained at point $A:=(r_w^L,\tilde{r}_m)$. Let the maximum value of the second function be $r_m^U$ and let it be attained at point $B:=(\tilde{r}_w,r_m^U)$. By property~(\ref{eq:property3}) of~$S$, $C:=(r_w^L,r_m^U)\in S$.
By way of contradiction, let us assume that~$C\not\in K$.
We know that $\{A,C\}\subset S$ but they are in different connected components of~$S$, so there should be a point on the interval~$(A,C)$  that does not belong to~$S$. In other words 
\[
\exists \lambda\in(0,1):(r_w^L,\lambda\tilde{r}_m+(1-\lambda)r_m^U)\not \in S_1\cap S_2.
\]
Because $\{A,C\}\subset S_1$ we have~$F^{-1}_{\Delta_w}(1-r_w^L)\leq g_w(r_w^L,\tilde{r}_m,1/r)$ and~$F^{-1}_{\Delta_w}(1-r_w^L)\leq g_w(r_w^L,r_m^U,1/r)$, which implies that for all~$\lambda\in(0,1)$, $F^{-1}_{\Delta_w}(1-r_w^L)\leq g_w(r_w^L,\lambda\tilde{r}_m+(1-\lambda)r_m^U,1/r)$, hence~$(r_w^L,\lambda\tilde{r}_m+(1-\lambda)r_m^U)\in S_1$.
So there should exist a~$\lambda\in(0,1)$ such that~$(r_w^L,\lambda\tilde{r}_m+(1-\lambda)r_m^U)\not\in S_2$, which implies
\[
F^{-1}_{\Delta_m}(1-\lambda\tilde{r}_m-(1-\lambda)r_m^U)< g_m(\lambda\tilde{r}_m+(1-\lambda)r_m^U,r_w^L,r).
\]
But $g_m$ is increasing in second variable, so for any~$r_w\geq r_w^L$,
\[
F^{-1}_{\Delta_m}(1-\lambda\tilde{r}_m-(1-\lambda)r_m^U)< g_m(\lambda\tilde{r}_m+(1-\lambda)r_m^U,r_w,r),
\] which implies that~$(r_w,\lambda\tilde{r}_m+(1-\lambda)r_m^U)$ is not in~$S_2$, hence not in~$K$.
But by definition, $r_w^L$ is the minimum value of the first coordinate that a point in~$K$ may have, so~$(r_w,\lambda\tilde{r}_m+(1-\lambda)r_m^U)$ doesn't belong to~$K$ for any~$r_w$.
However points~$A$ and~$B$ lie on different sides of this line, so they cannot be both in the connected set~$K$, which yields the desired contradiction.
\end{proof}


\begin{fact}\label{fact:upperleft2}
Let~$K$ be a connected component of~$S$ with an interior point~$(r_w^I,r_m^I)$. Then~$r_w^I>r_w^L$ and~$r_m^I<r_m^U$, where $(r_w^L,r_m^U)$ is the upperleft corner of~$K$.
\end{fact}


\begin{fact}\label{fact:delta}
    Let~$K$ be a connected component of~$S$ with upperleft corner~$(r_w^L,r_m^U)$ and another point~$(r_w^I,r_m^I)$ such that~$r_w^I>r_w^L$ and~$r_m^I<r_m^U$. Then there exists a~$\bar{\delta}>0$ such that for any $0<\delta<\bar{\delta}$, the following hold.
\begin{eqnarray*}    
    F^{-1}_{\Delta_w}(1-r_w^L-\delta) & \leq & g_w(r_w^L+\delta,r_m^U,1/r), \\
    F^{-1}_{\Delta_m}(1-r_m^U+\delta) & \geq & g_m(r_m^U-\delta,r_w^L,r).
    \end{eqnarray*}
\end{fact}

\begin{proof}[Proof of Fact~\ref{fact:delta}]
Let $\bar{\delta}=min(r_w^I-r_w^L,r_m^U-r_m^I)$, then for any $0<\delta<\bar{\delta}$, we have
$r_w^L<r_w^L+\delta<r_w^I $ and $r_m^I<r_m^U-\delta<r_m^U $.
By way of contradiction, assume first that there is a~$0<\delta<\bar{\delta}$ such that
\[
F^{-1}_{\Delta_w}(1-r_w^L-\delta)> g_w(r_w^L+\delta,r_m^U,1/r).
\]
Now, $g_w$ is increasing in its second variable. Hence, for all~$r_m\leq r_m^U$,
\[
F^{-1}_{\Delta_w}(1-r_w^L-\delta)> g_w(r_w^L+\delta,r_m,1/r),
\]
which implies~$(r_w^L+\delta,r_m)\not\in K$. Now~$r_m^U$ is the maximum in~$K$ of the second coordinate, so that~$(r_w^L+\delta,r_m)\not\in K$ is true for all~$r_m$.
However, this line separates $(r_w^L,r_m^U)$ and $(r_w^I,r_m^I)$ that lie in the same connected component~$K$, which yields a contradiction. The proof of the second inequality proceeds in the same way.
\end{proof}


\begin{fact}\label{fact:eitheror}
    Let~$K$ be a connected component of~$S$ with upperleft corner~$(r_w^L,r_m^U)$. let~$(r_w^I,r_m^I)$ be another point in~$K$ such that~$r_w^I>r_w^L$ and~$r_m^I<r_m^U$. Then the following hold.
    \begin{enumerate}
        \item If $r_w^L>0$ then
        \begin{itemize}
            \item \textbf{either} there exists such $\bar{\delta}>0$ that for any $0<\delta<\bar{\delta}$
             $$F^{-1}_{\Delta_w}(1-r_w^L+\delta)\geq g_w(r_w^L-\delta,r_m^U,1/r)$$
             \item \textbf{or} there exists such $\bar{\delta}>0$ that for any $0<\delta<\bar{\delta}$ function 
             $$W_{r_m^U}(r_w)=F^{-1}_{\Delta_w}(1-r_w)- g_w(r_w,r_m^U,1/r)$$
             changes sign infinitely many times on $r_w\in (r_w^L-\delta,r_w^L)$.
        \end{itemize} 
        \item If $r_m^U<1$ then
        \begin{itemize}
            \item \textbf{either} there exists such $\bar{\delta}>0$ that for any $0<\delta<\bar{\delta}$
             $$F^{-1}_{\Delta_m}(1-r_m^U-\delta)\leq g_m(r_m^U+\delta,r_w^L,r)$$
             \item \textbf{or} there exists such $\bar{\delta}>0$ that for any $0<\delta<\bar{\delta}$ function 
             $$M_{r_w^L}(r_m)=F^{-1}_{\Delta_m}(1-r_m)- g_m(r_m,r_w^L,r)$$
             changes sign infinitely many times on $r_m\in (r_m^U,r_m^U+\delta)$.
        \end{itemize} 
    \end{enumerate}
    \end{fact}

\begin{proof}[Proof of fact~\ref{fact:eitheror}]
We prove~(1). The proof of~(2) is identical. The two parts of our either-or statements cannot be true at the same time. So we just need to show that the two parts can not be false at the same time. Suppose the ``either'' statement in~(1) is false. Then, for~$\bar{\delta}=r_w^L$, there exists a~$\delta^{N_1}\in (0,\bar{\delta})$ such that
\[
F^{-1}_{\Delta_w}(1-r_w^L+\delta^{N_1})< g_w(r_w^L-\delta^{N_1},r_m^U,1/r).
\]
However, $(r_w^L,r_m^U)\in S_2$ and~$g_m$ is increasing in its second variable. Hence,
\[
F^{-1}_{\Delta_m}(1-r_m^U)\geq g_m(r_m^U,r_w^L-\delta^{N_1},r),
\]
which implies~$(r_w^L-\delta^{N_1},r_m^U)\in S$. Now~$(r_w^L,r_m^U)$ is the leftmost point in~$K$, so~$(r_w^L-\delta^{N_1},r_m^U)$ cannot be in~$K$. That means that the interval between these two points should contain a point that is not in~$S$; in other words, $\exists \delta^{P_1}\in (0, \delta^{N_1}):(r_w^L-\delta^{P_1},r_m^U)\not\in S$.
But~$(r_w^L,r_m^U)\in S_2$ and $g_m$ is increasing in second variable, hence
\[
F^{-1}_{\Delta_m}(1-r_m^U)\geq g_m(r_m^U,r_w^L-\delta^{P_1},r).
\]
So in order to have~$(r_w^L-\delta^{P_1},r_m^U)\not\in S$, we need~$(r_w^L-\delta^{P_1},r_m^U)\not\in S_1$, hence
\[
F^{-1}_{\Delta_w}(1-r_w^L+\delta^{P_1})> g_w(r_w^L-\delta^{P_1},r_m^U,1/r).
\]
Using the fact that the ``either statement in~(1) is false and repeating the same argument, we know that for~$\bar{\delta}=\delta^{P_1}$, there is a~$\delta^{N_2}\in (0,\delta^{P_1})$ and a~$\delta^{P_2}\in (0, \delta^{N_2})$, such that
\begin{eqnarray*}
F^{-1}_{\Delta_w}(1-r_w^L+\delta^{N_2}) & < & g_w(r_w^L-\delta^{N_2},r_m^U,1/r), \\
F^{-1}_{\Delta_w}(1-r_w^L+\delta^{P_2}) & > & g_w(r_w^L-\delta^{P_2},r_m^U,1/r).
\end{eqnarray*}
Repeating this process infinitely we get a sequence of positive numbers~$\delta^{N_1}>\delta^{P_1}>\delta^{N_2}>\delta^{P_2}>\delta^{N_3}>\delta^{P_3}>\cdots$ such that the function
\[
r_w \mapsto F^{-1}_{\Delta_w}(1-r_w)- g_w(r_w,r_m^U,1/r)
\]
is positive when~$r_w=r_w^L-\delta^{P_i}$ and negative when~$r_w=r_w^L-\delta^{N_i}$ for any natural~$i$. Hence the ``or'' part of~(1) is true, which completes the proof.
\end{proof}

\subsubsection{Existence}

By assumption, the earnings-only compositions satisfy~$0<r^e_w<r^e_m<1$. We also have~$F^{-1}_{\Delta_w}(1-r_w^e)< g_w(r_w^e,r_m^e,1/r)$ and~$F^{-1}_{\Delta_m}(1-r_m^e)> g_m(r_m^e,r_w^e,r)$. Therefore, by continuity of functions~$F^{-1}_{\Delta_t}$ and~$g_t$, $(r_w^e,r_m^e)$ is an interior point of~$S$. So there is a connected component~$K^e$ of~$S$ that contains $(r_w^e,r_m^e)$. Call~$(r_w^\ast,r_m^\ast)$ the upperleft corner of~$K^e$, as defined in fact~\ref{fact:upperleft1}. By fact~\ref{fact:upperleft2}, $(r_w^\ast,r_m^\ast)$ satisfies~$r_w^*<r_w^e<r_m^e<r_m^*$. Now by fact~\ref{fact:delta}, there exists a~$\bar{\delta}>0$, such that for any $0<\delta<\bar{\delta}$,
\begin{eqnarray*} 
F^{-1}_{\Delta_w}(1-r_w^*-\delta) & \leq & g_w(r_w^*+\delta,r_m^*,1/r), \\
F^{-1}_{\Delta_m}(1-r_m^*+\delta) & \geq & g_m(r_m^*-\delta,r_w^*,r).
\end{eqnarray*}
Given assumption~\ref{ass:reg2} the ``or'' parts of fact~\ref{fact:eitheror} cannot be true. Hence, there exists a~$\bar{\delta}>0$ (possibly different from the previous one) such that for any~$0<\delta<\bar{\delta}$ the following hold.
\begin{eqnarray*} 
F^{-1}_{\Delta_w}(1-r_w^*+\delta) & \geq & g_w(r_w^*-\delta,r_m^*,1/r), \\
F^{-1}_{\Delta_m}(1-r_m^*-\delta) & \leq & g_m(r_m^*+\delta,r_w^*,r).
\end{eqnarray*}
Hence~$(r_w^*,r_m^*)$ is a stable amplified equilibrium.

\subsubsection{Uniqueness}

For any~$\bar\varepsilon\in(0,1/2)$, set~$\varepsilon:=\min\left( \bar{\varepsilon},r_w^e,r_m^e-r_w^e,1-r_m^e\right)/4$. The functions~$h_w$ and~$h_m$ are continuous so there exist~$M_w$ and~$M_m$ such that 
\begin{eqnarray*}
\forall (r_w,r_m)\in [\varepsilon,1-\varepsilon]^2: M_w>\left|\tilde{g}_w(r_w,r_m,1/r)\right|\text{ and } M_m>\left|\tilde{g}_m(r_m,r_w,r)\right|,
\end{eqnarray*}
where~$\tilde g_t:=g_t/c_t$ is defined analogously to~$\tilde h_t$ for~$t\in\{w,m\}$.
The functions~$h_w$ and~$h_m$ are continuously differentiable so there exist~$D_w$ and~$D_m$ such that 
\begin{eqnarray*}
\begin{array}{l}
\forall (r_w,r_m)\in [\varepsilon,1-\varepsilon]^2: 4\left|\frac{\partial }{\partial r_w}\tilde{g}_w(r_w,r_m,1/r)\right|+4\left|\frac{\partial }{\partial r_m}\tilde{g}_w(r_w,r_m,1/r)\right| < D_w, \\
\forall (r_w,r_m)\in [\varepsilon,1-\varepsilon]^2:  4\left|\frac{\partial }{\partial r_m}\tilde{g}_m(r_m,r_w,r)\right|+4\left|\frac{\partial }{\partial r_w}\tilde{g}_m(r_m,r_w,r)\right| < D_m.
\end{array}
\end{eqnarray*}
The functions~$F_{\Delta_w}$ and~$F_{\Delta_m}$ are continuously differentiable, so there exist~$U_w$ and~$U_m$ such that
\begin{eqnarray*}
&&\forall x\in [F^{-1}_{\Delta_w}(1-r_w^e-\varepsilon),F^{-1}_{\Delta_w}(1-r_w^e+\varepsilon)]: U_w>F'_{\Delta_w}(x), \\
&&\forall x\in [F^{-1}_{\Delta_m}(1-r_m^e-\varepsilon),F^{-1}_{\Delta_m}(1-r_m^e+\varepsilon)]: U_m>F'_{\Delta_m}(x).
\end{eqnarray*}
Define
\begin{eqnarray*}
v:=\min\left(F^{-1}_{\Delta_w}(1-r_w^e+\varepsilon), -F^{-1}_{\Delta_w}(1-r_w^e-\varepsilon), F^{-1}_{\Delta_m}(1-r_m^e +\varepsilon), -F^{-1}_{\Delta_m}(1-r_m^e -\varepsilon) \right),
\end{eqnarray*}
and~$\bar{c}_t=\min\left(v/M_t,1/(D_tU_t) \right)$ for~$t\in\{w,m\}.$
Assume~$c_w\in(0,\bar c_w)$ and $c_m\in(0,\bar c_m)$, and let~$(r_w^*,r_m^*)$ be a an equilibrium with compositions in~$[\varepsilon,1-\varepsilon]^2$. Then the following equations must hold.
\begin{eqnarray*}
F^{-1}_{\Delta_w}(1-r_w^*) & = &  c_w\tilde{g}_w(r_w^*,r_m^*,1/r), \\
F^{-1}_{\Delta_m}(1-r_m^*) & = & c_m\tilde{g}_m(r_m^*,r_w^*,r).
\end{eqnarray*}
Because functions~$F^{-1}_{\Delta_w}$ and~$F^{-1}_{\Delta_m}$ are monotone and because~${c}_w<v/M_w$, and~${c}_m<v/M_m$, we have~$r_t^*\in(r_t^e-\varepsilon,r_t^e+\varepsilon)$ for~$t\in\{w,m\}$. Now, given that $r_w^e+\varepsilon<r_m^e-\varepsilon$ the equilibrium is non-contrarian, so it has to be amplified.
Now because~${c}_w<1/(D_wU_w)$ and~${c}_m<1/(D_mU_m)$, we have
\begin{eqnarray*}
\frac{\partial }{\partial r_w}\left(F^{-1}_{\Delta_w}(1-r_w^*) \right) & < & \frac{\partial }{\partial r_w}\left(c_w\tilde{g}_w(r_w^*,r_m^*,1/r)\right), \\
\frac{\partial }{\partial r_m}\left(F^{-1}_{\Delta_m}(1-r_m^*) \right) & < & \frac{\partial }{\partial r_m}\left(c_m\tilde{g}_m(r_m^*,r_w^*,r)\right),
\end{eqnarray*}
so $(r_w^*,r_m^*)$ is stable.
To prove uniqueness, apply the implicit function theorem to
\begin{eqnarray}\label{eq:imp1}
F^{-1}_{\Delta_w}(1-r_w) =  c_w\tilde{g}_w(r_w,r_m,1/r)
\end{eqnarray}
to obtain
\begin{eqnarray*}
\frac{\partial r_m }{\partial r_w} & = & \frac{\frac{1}{c_w}\frac{\partial }{\partial r_w}\left[ F^{-1}_{\Delta_w}(1-r_w) \right]-\frac{\partial }{\partial r_w}\left[ \tilde{g}_w(r_w,r_m,1/r) \right]}{\frac{\partial }{\partial r_m}\left[ \tilde{g}_w(r_w,r_m,1/r) \right]} \\
& < & \frac{\frac{1}{c_w}\frac{-1}{U_w}+D_w/4}{D_w/4} \; < \; \frac{\frac{-D_wU_w}{U_w}+D_w/4}{D_w/4} \; = \; -3
\end{eqnarray*}
Hence, the function~$r_m(r_w)$ defined implicitly from~(\ref{eq:imp1}) has slope lower than~$-3$. Similarly, the function~$r_w(r_m)$ defined implictly from
\begin{eqnarray*}
F^{-1}_{\Delta_m}(1-r_m) =  c_m\tilde{g}_m(r_m,r_w,r)
\end{eqnarray*}
has slope larger than~$-1/3$. These two curves cannot cross more than once, and equilibrium~$(r_w^*,r_m^*)$ is unique.
Finally, existence follows from the fact that curve~$r_m(r_w)$ is above point~$A:=(r_w^e-\varepsilon,r_m^e+\varepsilon)$ and below point~$B:=(r_w^e,r_m^e)$, whereas curve~$r_w(r_m)$ is below~$A$ and above~$B$, so they must intersect. This completes the proof of proposition~\ref{prop:ampli}.


\subsection{Proof of proposition~\ref{prop:sym}} This follows directly from definitions~\ref{def:eq} and~\ref{def:stable}.


\subsection{Proof of proposition~\ref{prop:1}}

\begin{proof}
We first show that an interior solution must satisfy the expressions given in the proposition, and must be an amplified equilibrium if \[\bar\gamma:=\max\{\gamma_wr_m^e/r_w^e+\gamma_m,\gamma_w+\gamma_m(1-r_w^e)/(1-r_m^e)\}<1.\]
An interior solution is a pair of composition $(r_w^*,r_m^*)$ that satisfies Equations (1.3) and (1.4). Under Assumption 2, these two equations become
    \begin{align*}
        C_w \frac{r_w^e - r_w^*}{[r_w^*(1-r_w^*)]^\beta} & = c_w \cdot \frac{1}{r} \cdot \frac{r_m^* - r_w^*}{r_w^*(1-r_w^*)} = c_w \cdot \left( \frac{\mu_{1}^*}{\mu_{1w}^*} - \frac{\mu_{2}^*}{\mu_{2w}^*} \right), \\
        C_m \frac{r_m^e - r_m^*}{[r_m^*(1-r_m^*)]^\beta} & = c_m \cdot r \cdot \frac{r_w^* - r_m^*}{r_m^*(1-r_m^*)} = c_m \cdot \left( \frac{\mu_{1}^*}{\mu_{1m}^*} - \frac{\mu_{2}^*}{\mu_{2m}^*} \right).
    \end{align*}
    If $\beta = 1$, then, applying the definitions of $\gamma_w$ and $\gamma_m$, these equations reduce to 
    \begin{align*}
        r_w^e - r_w^* & = \gamma_w \cdot (r_m^* - r_w^*), \\
        r_m^e - r_m^* & = \gamma_m \cdot (r_w^* - r_m^*),
    \end{align*}
    which imply that
    \begin{align}
        r_w^* = & r_w^e - \frac{\gamma_w}{1-\gamma_w - \gamma_m} (r_m^e - r_w^e)\label{eq:solutionwbeta=1}. \\
        r_m^* = & r_m^e + \frac{\gamma_m}{1-\gamma_w - \gamma_m} (r_m^e - r_w^e)\label{eq:solutionmbeta=1}.
    \end{align}
    Since $\bar\gamma<1$, we have $\gamma_m+\gamma_w<1$. \eqref{eq:solutionwbeta=1} and \eqref{eq:solutionmbeta=1} then imply that $r_w^*<r_w^e$ and $r_m^e<r_m^*$. Therefore, if an interior equilibrium exists, then it must be an amplified equilibrium. 

    Next, we show that if $\bar\gamma<1$, then an interior equilibrium exists. Simple algebra shows that \eqref{eq:solutionwbeta=1} and \eqref{eq:solutionmbeta=1} have an interior solution if and only if $\gamma_m < 1 - \max\{\frac{r_m^e}{r_w^e},\frac{1-r_m^e}{1-r_w^e}\} \gamma_w$ and $\gamma_w < 1 - \max\{\frac{r_w^e}{r_m^e},\frac{1-r_w^e}{1-r_m^e}\} \gamma_m$. Since $r_m^e > r_w^e$, these conditions are reduced to $\gamma_m < 1 - \frac{r_m^e}{r_w^e} 
        \gamma_w$ and $\gamma_m < \frac{1-r_m^e}{1-r_w^e} (1-\gamma_w)$, which is equivalent to $\bar\gamma<1$.

Finally, simple algebra shows that if $\bar\gamma<1$, then there is no corner equilibrium (nor partial-segregation equilibrium). 
\end{proof}


\subsection{Proof of proposition~\ref{prop:beta1}}

The following lemma gives a complete characterization of all equilibria, including unstable ones. Propositions~\ref{prop:beta1} and~\ref{prop:nudge} are straightforward corollaries.

\begin{lemma}\label{lem:17*}
    Suppose assumption~ 2 holds with~$\beta<1$ and the earnings-only compositions satisfy~$r^e_w<r^e_m$, then $(r_w^*,r_m^*)=(0,1)$ and $(r_w^*,r_m^*)=(1,0)$ are always equilibria and
    \begin{enumerate}
\item for any~$\bar\varepsilon\in(0,1/2)$, there is an $\varepsilon<\bar\varepsilon$ and a pair~$(\bar \gamma_w,\bar \gamma_m)$ such that for any~$\gamma_w\in(0,\bar \gamma_w)$ and~$\gamma_m\in(0,\bar \gamma_m)$, there are exactly 17 equilibria, 7 of which are stable. These equilibria are given in table~\ref{table:17*}, where
\begin{eqnarray*}
r_w^{(1)}<r_w^{(2)}<r_w^{(3)}<\varepsilon<r_w^{(4)}<r_w^{(5)}<r_w^{(6)}<r_w^e<r_w^{(7)}<r_w^{(8)}<1-\varepsilon<r_w^{(9)}<r_w^{(10)}<r_w^{(11)},\\
r_m^{(1)}<r_m^{(2)}<r_m^{(3)}<\varepsilon<r_m^{(4)}<r_m^{(5)}<r_m^e<r_m^{(6)}<r_m^{(7)}<r_m^{(8)}<1-\varepsilon<r_m^{(9)}<r_m^{(10)}<r_m^{(11)}.
\end{eqnarray*}

\item 
\begin{enumerate}
\item The only equilibrium in $(r_w^{(2)},r_w^{(10)})\times(r_m^{(2)},r_m^{(10)}) $ is \textbf{($\mathbf{r_w^{(6)},r_m^{(6)}}$)} and the two vertices of this square are unstable equilibria.
\item The vectors $(\dot{r}_w,\dot{r}_m)$ on the sides of the square $[r_w^{(2)},r_w^{(10)}]\times[r_m^{(2)},r_m^{(10)}] $ are directed inside it.
\end{enumerate} 
\item 
\begin{enumerate}
    \item $r_w^{(2)}$ is increasing in $\gamma_w$ and $r_m^{(2)}$  is increasing in $\gamma_m$.     
    \item $r_w^{(10)}$ is decreasing in $\gamma_w$ and $r_m^{(10)}$ is decreasing in $\gamma_m$.
\end{enumerate}

\item If, in addition, $\gamma_w=\gamma_m=\gamma$ and~$r_w^e=1-r_m^e$. Then
\begin{enumerate}
\item $r_w^{(2)}+r_m^{(10)}=1$ and $r_m^{(2)}+r_w^{(10)}=1$.
\item $r_w^{(2)}$ and $r_m^{(2)}$  are increasing in $\gamma$ and decreasing in $\beta$.
\item $r_w^{(10)}$ and $r_m^{(10)}$ are decreasing in $\gamma$ and increasing in $\beta$.
\end{enumerate} 
\end{enumerate}
\end{lemma}

\begin{table}
\begin{center}
\begin{tabular}{ m{8em}|m{8em}|m{8em}|m{8em} } 
   & $0\leq r_w<\varepsilon$&$\varepsilon\leq r_w\leq 1-\varepsilon$ & $1-\varepsilon<r_w\leq 1$ \\ \\ \hline \\
 $1-\varepsilon<r_m\leq 1$& \textbf{(0,1)},{(0,$r_m^{(9)}$)}, {($r_w^{(2)}$,$r_m^{(10)}$)},{($r_w^{(3)}$,$1$)} & \textbf{($\mathbf{r_w^{(4)}}$,1)},{($r_w^{(5)}$,$r_m^{(11)}$)} &  \\ \\ \hline \\
 $\varepsilon\leq r_m\leq 1-\varepsilon$& \textbf{(0,$\mathbf{r_m^{(8)}}$)},{($r_w^{(1)}$,$r_m^{(7)}$)} & \textbf{($\mathbf{r_w^{(6)},r_m^{(6)}}$)} & {($r_w^{(11)}$,$r_m^{(5)}$)}\textbf{(1,$\mathbf{r_m^{(4)}}$)} \\ \\ \hline \\
$0\leq r_m<\varepsilon$ & & {($r_w^{(7)}$,$r_m^{(1)}$)},\textbf{($\mathbf{r_w^{(8)}}$,0)} & 
{($r_w^{(9)}$,0)},{($r_w^{(10)}$,$r_m^{(2)}$)}, {(1,$r_m^{(3)}$)},\textbf{(1,0)} 
\\
\end{tabular}
\vskip5pt
\caption{Enumeration of all equilibria in case~$\beta<1$. Stable equilibria are in bold.}
\label{table:17*}
\end{center}
\end{table}

\begin{proof}[Proof of lemma~\ref{lem:17*}]
Consider the dynamic system:
\begin{eqnarray*}
\dot{r}_w  & = &  F^{-1}_{\Delta_w}(1-r_w) -  g_w(r_w,r_m,1/r) \\
\dot{r}_m & = &  F^{-1}_{\Delta_m}(1-r_m) - g_m(r_m,r_w,r).
\end{eqnarray*}

Using our parameterization, we can rewrite it as
\begin{eqnarray*}
\dot{r}_w & = & c_w \cdot \frac{1}{r} \cdot \frac{1}{r_w(1-r_w)}\left[  \frac{1}{\gamma_w}\frac{r_w^e-r_w}{[r_w(1-r_w)]^{\beta-1}}  + r_w  -  r_m\right],  \\ 
\dot{r}_m & = & c_m \cdot r \cdot \frac{1}{r_m(1-r_m)}\left[  \frac{1}{\gamma_m}\frac{r_m^e-r_m}{[r_m(1-r_m)]^{\beta-1}}  + r_m -  r_w\right].
\end{eqnarray*}

Define the function
\begin{eqnarray*}
d(x;r^e,\gamma,\beta) & = & \frac{1}{\gamma_t}\frac{r^e-x}{[x(1-x)]^{\beta-1}}  + x \\
& = & \frac{1}{\gamma}(r^e-x)\left[\frac{1}{4}-\left(\frac{1}{2}-x\right)^2\right]^{1-\beta} + x.
\end{eqnarray*}
Notice that $d(r_w,r_w^e,\gamma_w,\beta)-r_m$ always has the same sign as $F^{-1}_{\Delta_w}(1-r_w) -  g_w(r_w,r_m,r)$. The same is true for $d(r_m,r_m^e,\gamma_m,\beta)-r_w$. So the position of equilibria can be characterized by the function~$d$.
The derivative~$d'$ of function~$d$ with respect to its variable~$x$ can be written in the following ways.
\begin{eqnarray}
&& d'(x;r^e,\gamma,\beta) \nonumber \\ 
&& \hskip25pt = \frac{1}{\gamma}\cdot \frac{(1-\beta)r^e(1-x)+(1-\beta)(1-r^e)x-(3-2\beta)x(1-x)}{\left[x(1-x) \right]^{\beta}}+1 \hskip25pt \label{eq:17-1} \\
&& \hskip25pt = \frac{1}{\gamma}\cdot \frac{(r^e-x)\left[ 0.5-r^e +(3-2\beta)(0.5-x)     \right] -r^e(1-r^e)}{\left[x(1-x) \right]^{\beta}}+1. \hskip25pt \label{eq:17-2}
\end{eqnarray}
From expression~(\ref{eq:17-1}), we deduce that the function $d(r_w;r_w^e,\gamma_w,\beta)$ is increasing on $r_w\in\left[0,r_w^e(1-r_w^e)(1-\beta)/4\right]$ and on~$r_w\in[1-r_w^e(1-r_w^e)(1-\beta)/4,1]$, with derivative strictly more than~$1$. From expression~(\ref{eq:17-2}), we deduce that if $\gamma_w$ is such that $\gamma_w<r_w^e(1-r_w^e)/4$, then the function $d(r_w;r_w^e,\gamma_w,\beta)$ is decreasing on $r_w\in\left[r_w^e-r_w^e(1-r_w^e)/4,r_w^e+r_w^e(1-r_w^e)/4\right]$ with derivative strictly less than~$-1$.
\end{proof}

We now find the upper bounds~$(\bar \gamma_w,\bar \gamma_m)$.
For an arbitrary~$\bar\varepsilon\in(0,1/2)$ as in the statement of the proposition, define
\begin{eqnarray*}
\varepsilon=\min\left( \bar{\varepsilon},\frac{r_w^e(1-r_w^e)}{4}(1-\beta),\frac{r_m^e(1-r_m^e)}{4}(1-\beta)\right).
\end{eqnarray*}
The function $\frac{r_w^e-r_w}{[r_w(1-r_w)]^{\beta-1}}$ is continuous and positive on
$$\left[\frac{r_w^e(1-r_w^e)}{4}(1-\beta),r_w^e -\frac{r_w^e(1-r_w^e)}{4}(1-\beta)\right],$$ 
and it is continuous and negative on 
$$\left[r_w^e+\frac{r_w^e(1-r_w^e)}{4}(1-\beta),1-\frac{r_w^e(1-r_w^e)}{4}(1-\beta)\right].$$
So by maximum value theorem there exist~$l_1^w>0$ and~$l_2^w>0$ such that 
\begin{eqnarray*}
\begin{array}{l}
\forall r_w \in \left[\varepsilon,r_w^e -\frac{r_w^e(1-r_w^e)}{4}(1-\beta)\right]: l^w_1 \leq \frac{r_w^e-r_w}{[r_w(1-r_w)]^{\beta-1}}, \\ 
\forall r_w \in \left[r_w^e+\frac{r_w^e(1-r_w^e)}{4}(1-\beta),1-\varepsilon\right]: l^w_2 \leq \left|\frac{r_w^e-r_w}{[r_w(1-r_w)]^{\beta-1}}\right|.
\end{array}
\end{eqnarray*}
Defining
\begin{eqnarray*}
\bar{\gamma}_w=\min\left\{\frac{r_w^e(1-r_w^e)}{4},l^w_1,l^w_2  \right\},
\end{eqnarray*}
we can show that for all~$\gamma_w\in(0,\bar\gamma_w)$, we have~$d(r_w;r_w^e,\gamma_w,\beta)>1$ on~$[\varepsilon,r_w^e -r_w^e(1-r_w^e)(1-\beta)/4]$ and~$d(r_w;r_w^e,\gamma_w,\beta)<0$ on~$[r_w^e+r_w^e(1-r_w^e)(1-\beta)/4,1-\varepsilon]$.

Now we can collect all the facts that we proved about function~$d_w(\cdot):=d(\cdot;r_w^e,\gamma_w,\beta)$ for any~$\gamma_w\in(0,\bar\gamma_w)$. 
\begin{itemize}
    \item On $[0,\varepsilon]$ $d_w$ is strictly increasing from $0$ to $d_w(\varepsilon)>1$ with derivative strictly larger than 1.
    \item On $\left[\varepsilon,r_w^e -r_w^e(1-r_w^e)(1-\beta)/4\right]$, $d_w$ is strictly larger than $1$.
    \item On $\left[r_w^e -r_w^e(1-r_w^e)(1-\beta)/4,r_w^e +r_w^e(1-r_w^e)(1-\beta)/4\right]$ is strictly decreasing from 
    $d_w\left(r_w^e -r_w^e(1-r_w^e)(1-\beta)/4\right)>1$
    to
    $d_w\left(r_w^e +r_w^e(1-r_w^e)(1-\beta)/4\right)<0$
    with derivative strictly smaller than -1.
    \item On $\left[r_w^e+r_w^e(1-r_w^e)(1-\beta)/4,1-\varepsilon\right]$ $d_w$ is strictly smaller than 0.
    \item On $[1-\varepsilon,1]$, $d_w$ is strictly increasing from $d_w(1-\varepsilon)<0$ to $1$ with derivative strictly larger than 1.
\end{itemize}

Given that our model is symmetrical,  for any $\gamma_m\in(0,\bar{\gamma}_m)$, where~$\bar\gamma_m$ is defined analogously to~$\bar\gamma_w$, the function~$d_m(\cdot):=d(\cdot;r_m^e,\gamma_m,\beta)$ would have the same behaviour.
We can now characterize the full set of equilibria. First, we look at the vertices and we can check directly that~$(0,1)$ and~$(1,0)$ are stable equilibria, whereas~$(0,0)$ and~$(1,1)$ are not equilibria.
Then, all intersections~$(r_w^*,r_m^*)$ of~$d_w$ and~$d_m$ with each other or with the boundary of~$[0,1]^2$ minus its vertices are equilibria. Such equilibria are stable if~$d_w'(r_w^*)<0$ and~$d_m'(r_m^*)<0$, and unstable if~$d_w'(r_w^*)>0$ or~$d_m'(r_m^*)>0$. This completes the proof of~(1) and~(2)(a) in lemma~\ref{lem:17*}.

We now prove~(2)(b) of lemma~\ref{lem:17*}. Since~$(r_w^{(2)},r_m^{(10)})$ is an equilibrium, we have
\begin{eqnarray*}
\begin{array}{lllll}
        \dot{r}_w(r_w^{(2)},r_m^{(10)}) & = & 0 & = & F^{-1}_{\Delta_w}(1-r_w^{(2)}) -  g_w(r_w^{(2)},r_m^{(10)},1/r), \\
        \dot{r}_m(r_w^{(2)},r_m^{(10)}) & = & 0 & = & F^{-1}_{\Delta_m}(1-r_m^{(10)}) - g_m(r_m^{(10)},r_w^{(2)},r).
\end{array}
\end{eqnarray*}  
Given monotonicity of~$g_w$ and~$g_m$, we have
\begin{eqnarray*}
\begin{array}{l}
\forall r_m<r_m^{(10)}:\dot{r}_w(r_w^{(2)},r_m)>0, \mbox{ and }
\forall r_w>r_w^{(2)}:\dot{r}_m(r_w,r_m^{(10)})<0.
\end{array}
\end{eqnarray*}
Analogous inequalities hold relative to equilibrium~$(r_w^{(10)},r_m^{(2)})$, and the result follows. Parts~(3) and~(4)(b) and~(c) are proved by the implicit function theorem and inspection of the partial derivatives. There remains to prove part~(4)(a). By way of contradiction, assume that~$r_w^{(2)}\neq 1-r_m^{(10)}$. By symmetry, if~$(a,b)$ is a solution to the system of equations, then~$(1-b,1-a)$ is also a solution. So both~$(r_w^{(2)},r_m^{(10)})$ and~$(1-r_m^{(10)}),1-r_w^{(2)})$ are equilibria. But both of this points lie in~$(0,\varepsilon)\times (1-\varepsilon,1)$ and we have shown that~$d_w$ and~$d_m$ can have only one intersection in this area. This yields the desired contradiction and completes the proof of lemma~\ref{lem:17*}.


\subsection{Proof of proposition~\ref{prop:1beta}}
\hfill\\

\noindent{\bf Proof of uniqueness of a stable amplified equilibrium:}

We already have a general result that there is a stable amplified equilibrium in $[\varepsilon,1-\varepsilon]^2$. So to complete this proof we would show that there are a ~$0<\bar \gamma_w,0<\bar \gamma_m$ and  $0<\varepsilon_w,0<\varepsilon_m$ such that for any~$\gamma_w\in(0,\bar \gamma_w)$ and~$\gamma_m\in(0,\bar \gamma_m)$ there are no equilibria outside of $[\varepsilon_w,1-\varepsilon_w]\times [\varepsilon_m,1-\varepsilon_m]$.
Set
\begin{eqnarray*}
\begin{array}{llll}
\bar{\gamma}_w=\frac{r_w^e(1-r_w^e)}{2}, & \bar{\gamma}_m=\frac{r_m^e(1-r_m^e)}{2}, &
\varepsilon_w=\frac{r_w^e(1-r_w^e)}{2}, & \varepsilon_m=\frac{r_m^e(1-r_m^e)}{2}.
\end{array}
\end{eqnarray*}
We know that 
\begin{eqnarray*}
F^{-1}_{\Delta_w}(1-r_w) -  g_w(r_w,r_m,1/r) & = & c_w \cdot \frac{1}{r} \cdot \frac{1}{r_w(1-r_w)}\left[  \frac{1}{\gamma_w}\frac{r_w^e-r_w}{[r_w(1-r_w)]^{\beta-1}}  + r_w  -  r_m\right].
\end{eqnarray*}
    Consider 
    \begin{eqnarray*}
    d_w(r_w) & = & \frac{1}{\gamma_w}\frac{r_w^e-r_w}{[r_w(1-r_w)]^{\beta-1}}  + r_w.
    \end{eqnarray*}
    
    We first show that~$\gamma_w<\bar{\gamma}_w$ and $r_w\leq\varepsilon_w$ imply~$d_w(r_w)>1$.
   Indeed, given that $\varepsilon_w<r_w^e$ we have
        \begin{eqnarray*}
        \frac{1}{\gamma_w}\frac{r_w^e-r_w}{[r_w(1-r_w)]^{\beta-1}}  + r_w \geq \frac{1}{\gamma_w}(r_w^e-r_w)\geq \frac{1}{\gamma_w}\left(r_w^e-\frac{r_w^e(1-r_w^e)}{2}\right)> \\
        \frac{1}{\gamma_w}\frac{r_w^e}{2}>\frac{2}{r_w^e(1-r_w^e)}\frac{r_w^e}{2}=\frac{1}{1-r_w^e}>1.
        \end{eqnarray*}
        So if $\gamma_w<\bar{\gamma}_w$ then there can be no equilibria with $r_w\leq\varepsilon_w$. Analogously if $\gamma_m<\bar{\gamma}_m$ then there can be no equilibria with $r_m\leq\varepsilon_m$. 
   
   We now show that~$\gamma_w<\bar{\gamma}_w$ and $r_w\geq 1-\varepsilon_w$ imply $d_w(r_w)<0$.
   Indeed, given that $1-\varepsilon_w>r_w^e$ we have
        \begin{eqnarray*}
        \frac{1}{\gamma_w}\frac{r_w^e-r_w}{[r_w(1-r_w)]^{\beta-1}}  + r_w \leq \frac{1}{\gamma_w}(r_w^e-r_w)+1\leq \frac{1}{\gamma_w}\left(r_w^e+\frac{r_w^e(1-r_w^e)}{2}-1\right)+1< \\
        \frac{1}{\gamma_w}\left(\frac{(1-r_w^e)}{2}-(1-r_w^e)\right)+1=\frac{-1}{\gamma_w}\frac{1-r_w^e}{2}+1\leq \frac{-2}{r_w^e(1-r_w^e)}\frac{1-r_w^e}{2}+1=1-\frac{1}{r_w^e}<0.
        \end{eqnarray*}
        So if $\gamma_w<\bar{\gamma}_w$ then there can be no equilibrium with $r_w\geq 1-\varepsilon_w$. Analogously if $\gamma_m<\bar{\gamma}_m$ then there can be no equilibrium with $r_m\geq 1-\varepsilon_m$. 
        Hence for any~$\gamma_w\in(0,\bar \gamma_w)$ and~$\gamma_m\in(0,\bar \gamma_m)$ there are no equilibria outside of $[\varepsilon_w,1-\varepsilon_w]\times [\varepsilon_m,1-\varepsilon_m]$.

\vskip15pt
\noindent{\bf Proof of Parts (1) and (2) under assumptions $\gamma_w=\gamma_m$ and $r_w^e+r_m^e=1$:}

To analyze the cases of $\beta\neq1$, let $\tilde\gamma_w(r_w):=\gamma_w(r_w(1-r_w))^{\beta-1}$ and $\tilde\gamma_m(r_w):=\gamma_m(r_m(1-r_m))^{\beta-1}$. For brevity, we sometimes omit the argument of $\tilde\gamma_w$ and $\tilde\gamma_m$.

Define two functions $\Gamma_w$ and $\Gamma_m$ by
\begin{align}
    \Gamma_w(r_w^e,\gamma_w,r_w,r_m)&:= r_w^e - r_w-\tilde\gamma_w(r_w)(r_m - r_w),\label{eq:lawofmotiongeneralw}\\
    \Gamma_m(r_w^e,\gamma_w,r_w,r_m)&:=r_m^e - r_m-\tilde\gamma_m(r_m)(r_w - r_m).\label{eq:lawofmotiongeneralm}
\end{align}
An interior equilibrium is a pair $(r_w,r_m)\in(0,1)^2$ that solves $\Gamma_w=0$ and $\Gamma_m=0$. That is, an interior equilibrium $(r_w,r_m)$ solves
        \begin{align}
        r_w = & r_w^e - \frac{\tilde\gamma_w}{1-\tilde\gamma_w - \tilde\gamma_m} (r_m^e - r_w^e),\label{eq:solutionwbetaall} \\
        r_m = & r_m^e + \frac{\tilde\gamma_m}{1-\tilde\gamma_w - \tilde\gamma_m} (r_m^e - r_w^e).\label{eq:solutionmbetaall}
        \end{align}
Subtracting \eqref{eq:solutionwbetaall} from \eqref{eq:solutionmbetaall}, we have, 
    \begin{align}
        r_m^*-r_w^* = & \frac{1}{1-\tilde\gamma_w - \tilde\gamma_m} (r_m^e - r_w^e)\label{eq:solutiondifferencebeta=1}.
    \end{align}

Moreover, in a dynamical system, for each group $t\in\{w,m\}$, $\Gamma_t$ and $\dot{r}_t$ share the same sign. Assume that $(r_w^e,r_m^e)\in(0,1)^2$, then obviously, when $\beta>1$, there is no corner equilibria, because, for example, if $r_w=0$ ($r_w=1$), then $\tilde\gamma_w(r_w)=0$ and thus $\dot{r}_w>0$ ($\dot{r}_w<0$), implying the system is not in equilibrium . Thus, in the rest of this proof, equilibrium refers to interior equilibrium.

    \textit{Step 1. We show that there is a unique amplified equilibrium (and it is stable). }

    Suppose by contradiction that there are two interior equilibria, denoted as $(r_w^1,r_m^1)$ and $(r_w^2,r_m^2)$, that have an amplification effect. Suppose that $r_m^2>r_m^1>r_m^e$. Since $r_w^e<1/2<r_m^e$, we have $\tilde\gamma(r_m^2)<\tilde\gamma(r_m^1)$. Moreover, both solutions satisfy \eqref{eq:solutionwbetaall} and \eqref{eq:solutionmbetaall}. 
    
If $\tilde\gamma(r_w^2)+\tilde\gamma(r_m^2)\leq\tilde\gamma(r_w^1)+\tilde\gamma(r_m^1)$, then \eqref{eq:solutionmbetaall} implies that $r_m^2<r_m^1$, a contradiction. 

If $\tilde\gamma(r_w^2)+\tilde\gamma(r_m^2)>\tilde\gamma(r_w^1)+\tilde\gamma(r_m^1)$, which is possible only if $\tilde\gamma(r_w^2)>\tilde\gamma(r_w^1)$, \eqref{eq:solutionwbetaall} implies that $r_w^2<r_w^1$. But since $r_w^1<r_w^e<1/2$, we have $\tilde\gamma(r_w^2)<\tilde\gamma(r_w^1)$, a contradiction. 

Combining the existence of a stable amplified equilibrium in Proposition \ref{prop:ampli}, we conclude that there is a unique amplified equilibrium and it is stable. 

    \textit{Step 2. We show that if $(r_w,r_m)$ is a solution, then $r_m=1-r_w$.}
    
    Suppose by contradiction that $(r_w,r_m)$ is a solution and $r_m\neq1-r_w$. Then,  $(1-r_m,1-r_w)$ is also a solution because it solves $\Gamma_w=0$ and $\Gamma_m=0$, where $\Gamma_w$ and $\Gamma_m$ are defined by \eqref{eq:lawofmotiongeneralw} and \eqref{eq:lawofmotiongeneralm}. 

    Since $r_m\neq1-r_w$, 
    either $(r_w,r_m)$ or $(1-r_m,1-r_w)$ is a contrarian equilibrium, by Step 1. Without loss of generality, assume that $(r_w,r_m)$ is a contrarian equilibrium, i.e., $r_m<r_w$. 
    $(r_w,r_m)$ must solve \eqref{eq:solutionwbetaall} and \eqref{eq:solutionmbetaall} (with $r_w^*$ and $r_m^*$ replaced by $r_w$ and $r_m$). Adding up these two equations, we have
    \begin{align}
        r_w+r_m=1+\frac{\tilde\gamma_w-\tilde\gamma_m}{\tilde\gamma_w + \tilde\gamma_m-1} (r_m^e - r_w^e).\label{eq:sumofsolutoin}
    \end{align}
    Since $(r_w,r_m)$ is a contrarian equilibrium, $(r_m-r_w)$ and $(r_m^e-r_w^e)$ have opposite signs, and thus by \eqref{eq:solutiondifferencebeta=1}, we have $\tilde\gamma_w + \tilde\gamma_m-1>0$. 
    
    Suppose first that $r_m>1-r_w$. And since $r_w>r_m$ (because the equilibrium is contrarian), we have $r_w>1-r_w$, i.e., $r_w>1/2$. We now show that $|r_m-1/2|<r_w-1/2$. First, by $r_m>1-r_w$, we have $1/2-r_m<r_w-1/2$; second, since $r_m<r_w$, we have $r_m-1/2<r_w-1/2$. Thus, $|r_m-1/2|<r_w-1/2$ holds. This implies that $(r_w(1-r_w))^{\beta-1}<(r_m(1-r_m))^{\beta-1}$ because the function $x(1-x)$ is quadratic and maximized at $x=1/2$. Applying our definition that $\tilde\gamma_w(r_w):=\gamma_w(r_w(1-r_w))^{\beta-1}$ and $\tilde\gamma_m(r_w):=\gamma_m(r_m(1-r_m))^{\beta-1}$, and the assumption that $\gamma_w=\gamma_m$, we have that $\tilde\gamma_w<\tilde\gamma_m$. Applying this inequality and $\tilde\gamma_w + \tilde\gamma_m-1>0$ (proved in the previous paragraph) to \eqref{eq:sumofsolutoin}, we have that $r_w+r_m<1$, which contradicts our assumption (at the beginning of this paragraph) $r_m>1-r_w$.  
    
    Similarly, if $r_m<1-r_w$, we also reach a contradiction.

    Therefore, if $(r_w,r_m)$ is a solution, then $r_m=1-r_w$.

    \textit{Step 3. We show that if $\beta'$ is such that there is no contrarian equilibrium, then for all $\beta>\beta'$, there is no contrarian equilibrium and thus the only equilibrium is an amplified equilibrium.}

From Step 2 and the fact that any interior equilibrium must solve \eqref{eq:solutiondifferencebeta=1}, any contrarian equilibrium must be of the form $(r_w,1-r_w)$, with $r_w>1/2$, that solves
    \begin{equation}
        1-2r_w =\frac{1}{1-2\tilde\gamma_w} (r_m^e - r_w^e),\label{eq:solutionwbetaallsymmetrydiffer}
    \end{equation}
    which is obtained from \eqref{eq:solutiondifferencebeta=1} by equating $\tilde\gamma_w$ and $\tilde\gamma_m$. 
For convenience, we rewrite \eqref{eq:solutionwbetaallsymmetrydiffer} as 
    \begin{equation}
        2r_w-1 =\frac{1}{2\tilde\gamma_w-1} (r_m^e - r_w^e).\label{eq:solutionwbetaallsymmetrydiffercontra}
    \end{equation}
    For $r_w>1/2$ and \eqref{eq:solutionwbetaallsymmetrydiffercontra} to hold, a contrarian equilibrium must satisfy $\tilde\gamma(r_w)>1/2$, i.e., a contrarian equilibrium, if it exists, is in the set $\{r_w\in[1/2,1]:2\gamma_w(r_w(1-r_w))^{\beta-1}>1\}$. Since for all $r_w\in[1/2,1)$, $(r_w(1-r_w))^{\beta-1}$ is strictly decreasing in $\beta$, we have, 
    \[\{r_w\in[1/2,1]:2\gamma_w(r_w(1-r_w))^{\beta-1}>1\}\subset \{r_w\in[1/2,1]:2\gamma_w(r_w(1-r_w))^{\beta'-1}>1\}\]
    for all $\beta>\beta'$. That is, for all $\beta>\beta'$, a contrarian equilibrium (under parameter $\beta$) must be in the set $\{r_w\in[1/2,1]:2\gamma_w(r_w(1-r_w))^{\beta'-1}>1\}$. We now show that if there is no contrarian equilibrium under tail parameter $\beta'$, then the LHS of \eqref{eq:solutionwbetaallsymmetrydiffercontra} is always strictly below the RHS of \eqref{eq:solutionwbetaallsymmetrydiffercontra} for all $r_w$ in the set $\{r_w\in[1/2,1]:2\gamma_w(r_w(1-r_w))^{\beta-1}>1\}$, which implies that there is no contrarian equilibrium (under tail parameter $\beta$).
    
    For any $\beta>\beta'$, if $r_w=1/2$ then the LHS of \eqref{eq:solutionwbetaallsymmetrydiffercontra} is 0, which is strictly below the RHS. Since under tail parameter $\beta'$, there is no contrarian equilibrium, it must be that for all $r_w$ such that $\tilde\gamma(r_w)>1/2$, the LHS of \eqref{eq:solutionwbetaallsymmetrydiffercontra} is strictly below the RHS; that is (applying the definition of $\tilde\gamma_w$),
    \[2r_w-1 <\frac{1}{2\gamma_w(r_w(1-r_w))^{\beta'-1}-1} (r_m^e - r_w^e).\]
    Since $\frac{1}{2\gamma_w(r_w(1-r_w))^{\beta'-1}-1}$ is strictly increasing in $\beta$ as long as $2\gamma_w(r_w(1-r_w))^{\beta'-1}>1$, we have, for all $\beta>\beta'$, 
    \[2r_w-1 <\frac{1}{2\gamma_w(r_w(1-r_w))^{\beta-1}-1} (r_m^e - r_w^e).\]
    Therefore, the LHS  of \eqref{eq:solutionwbetaallsymmetrydiffercontra} is strictly lower than the RHS  of \eqref{eq:solutionwbetaallsymmetrydiffercontra} for all $r_w$ in the set $\{r_w\in[1/2,1]:2\gamma_w(r_w(1-r_w))^{\beta-1}>1\}$, implying that there is no contrarian equilibrium under tail parameter $\beta>\beta'$. 
    
    \textit{Step 4. We show that there are at most two contrarian equilibria.}

   From previous steps, $(r_w,1-r_w)$ is a contrarian equilibrium if $r_w$ solves \eqref{eq:solutionwbetaall} and $r_w\in[1/2,1]$. We now assume $r_w\in[1/2,1]$ since we are analyzing contrarian equilibria. Equating $\tilde\gamma_w$ and $\tilde\gamma_m$ in \eqref{eq:solutionwbetaall}, we obtain
    \begin{equation}
        r_w =  r_w^e + \frac{\tilde\gamma_w}{2\tilde\gamma_w-1} (r_m^e - r_w^e).\label{eq:solutionwbetaallsymmetry}
    \end{equation}
    Rearranging \eqref{eq:solutionwbetaallsymmetry}, we have,
    \begin{equation}
       \tilde\gamma_w(2(r_w-r_w^e)-(r_m^e-r_w^e))= r_w - r_w^e,
    \end{equation}
    which, by applying $r_m^e+r_w^e=1$, is equivalent to
    \begin{equation}\label{eq:solutiontosymmetricen}
       \tilde\gamma_w(2r_w-1)= r_w - r_w^e.
    \end{equation}
Using the expression of $\tilde\gamma_w$, this equation is equivalent to
\begin{equation}\label{eq:solequationforbeta>1}
    G(r_w,\beta,\gamma_w):=\gamma_w (r_w(1-r_w))^{\beta-1}(2r_w-1)-r_w+r_w^e=0
\end{equation}
That is, $(r_w,1-r_w)$ is a contrarian equilibrium if $r_w\in[1/2,1]$ solves $G(r_w,\beta,\gamma_w)=0$. Note that at $r_w=1/2$, $G(r_w,\beta,\gamma_w)=-1/2+r_w^e<0$, and thus $(1/2,1/2)$ is not an equilibrium. So to look for contrarian equilibria, we now assume $r_w\in(1/2,1]$.
Let $f(r_w,\beta):=-1-2r_w+2r_w^2+(\beta-1)(2r_w-1)^2$. It can be shown that there is some function $\tilde f$ with $\tilde f(r_w)>0$ such that
\begin{equation}
    \frac{\partial^2G(r_w,\beta,\gamma_w)}{\partial r_w^2}=(\beta-1)\gamma_w\tilde f(r_w)f(r_w,\beta).
\end{equation}
We have 
\begin{equation*}
    \frac{\partial f(r_w,\beta)}{\partial r_w}=2(2r_w-1)+4(\beta-1)(2r_w-1)>0
\end{equation*}
since $\beta>1$ and $r_w>1/2$. From these curvature properties of $G$, there can be at most 3 solutions to \eqref{eq:solequationforbeta>1} for $r_w\in[1/2,1]$. And since 
    $G(1/2,\beta,\gamma_w)=-1/2+r_w^e<0$ and $G(1,\beta,\gamma_w)=-1+r_w^e<0$, the equation $G(r_w,\beta,\gamma_w)=0$ must have an even number of solutions in $[1/2,1]$. The fact that there are at most 3 solutions implies that there are at most two contrarian solutions. 

\textit{Step 5. If $\gamma_w<1-r_w^e$, then contrarian equilibrium does not exist. If $\gamma_w>1-r_w^e$ and $\beta$ is high, then a contrarian equilibrium exists.}

We first show that for any fixed $\beta>1$, if there is a contrarian when $\gamma_w=\gamma_w'$, there are two contrarian equilibria for all $\gamma_w>\gamma_w'$. Recall from Step 5 that $(r_w,1-r_w)$ is a contrarian equilibrium if $r_w\in[1/2,1]$ solves $G(r_w,\beta,\gamma_w)=0$. 
    Since  for all $\gamma_w>0$, $G(1/2,\beta,\gamma_w)=-1/2+r_w^e<0$ and $G(1,\beta,\gamma_w)=-1+r_w^e<0$, and  $G$ is strictly increasing in $\gamma_w$ for $r_w\in(1/2,1)$, if for some $\gamma_w'$ there is at least one solution to $G(r_w,\beta,\gamma_w')=0$ in $[1/2,1]$, then there are two solutions to $G(r_w,\beta,\gamma_w)=0$ in $[1/2,1]$ for all $\gamma_w>\gamma_w'$.

    If $\beta=1$ and $\gamma_w=1-r_w^e$, then $G(r_w,1,\gamma_w)<0$ for all $r_w\in[1/2,1)$ and $G(1,1,\gamma_w)=0$. Given the monotonicity of $G$ in $\gamma_w$ and $\beta$, we have, (a) if $\gamma_w\leq1-r_w^e$, then for all $\beta>1$, $G(r_w,\beta,\gamma_w)<0$ for all $r_w\in[1/2,1)$; and (b) if $\gamma_w>1-r_w^e$, then there is some $r_w'\in[1/2,1)$ such that $G(r_w',1,\gamma_w)>0$. For this $r_w'$, if $\beta$ is greater than but sufficiently close to 1, then $G(r_w',\beta,\gamma_w)>0$. And since $G(1/2,\beta,\gamma_w)>0$, there is an $r_w\in(1/2,r_w')$ such that $G(r_w,\beta,\gamma_w)=0$. The desired result follows.

\end{appendix}



\bibliography{Tipping}
\bibliographystyle{abbrvnat}

\end{document}